\documentclass[journal]{IEEEtran}
\usepackage{cite}
\usepackage[dvips]{graphicx} \graphicspath{{./}}
\DeclareGraphicsExtensions{.eps}

\usepackage[cmex10]{amsmath}
\usepackage{array}
\ifCLASSOPTIONcompsoc
  \usepackage[caption=false,font=normalsize,labelfont=sf,textfont=sf]{subfig}
\else
  \usepackage[caption=false,font=footnotesize]{subfig}
\fi
\usepackage{dblfloatfix}
\let\MYorigsubfloat\subfloat
\renewcommand{\subfloat}[2][\relax]{\MYorigsubfloat[]{#2}}
\usepackage{url}
\usepackage[english]{babel}
\usepackage{mathtools}
\usepackage{blindtext}
\usepackage{psfrag}
\usepackage{pstricks, pst-node, pst-plot, pst-circ}
\usepackage{moredefs}
\usepackage{color}
\usepackage{booktabs}
\usepackage{cleveref}
\usepackage{xifthen}
\usepackage{multirow}
\usepackage{tabu}
\usepackage{siunitx}
\usepackage{mathtools}

\usepackage{amssymb}
\usepackage{amsthm}
\usepackage{accents}
\usepackage{algpseudocode}
\usepackage{mathrsfs}
\usepackage{scrextend}

\newcommand{\ra}[1]{\renewcommand{\arraystretch}{#1}}

\makeatletter
\newcommand{\dontusepackage}[2][]{%
  \@namedef{ver@#2.sty}{9999/12/31}%
  \@namedef{opt@#2.sty}{#1}}
\makeatother

\usepackage{accents}
\usepackage{mathtools}

\DeclarePairedDelimiter\abs{\lvert}{\rvert}%
\DeclarePairedDelimiter\norm{\lVert}{\rVert}%

\makeatletter
\let\oldabs\abs
\def\abs{\@ifstar{\oldabs}{\oldabs*}}
\let\oldnorm\norm
\def\norm{\@ifstar{\oldnorm}{\oldnorm*}}

\newcommand\by[1]{\mathbf{r}_{#1}}

\newcommand\bbf[1]{{e}_{#1}}
\newcommand\bn[1]{\mathbf{n}_{#1}}
\newcommand\bxh[1]{\hat{\bx{}}_{#1}}
\newcommand\bx[1]{\mathbf{u}_{#1}}

\newcommand{\alp}[2]{%
\ifthenelse{\equal{#1}{}}{\ifthenelse{\equal{#2}{}}{\alpha}{\alpha^{(#2)}}}{\ifthenelse{\equal{#2}{}}{\alpha_{#1}}{\alpha^{(#2)}_{#1}}}}

\newcommand{\alpi}[1]{%
\ifthenelse{\equal{#1}{}}{\innov{\alpv{}}}{\innov{\alpha}_{#1}}}

\newcommand{\aln}[1]{%
\ifthenelse{\equal{#1}{}}{\bar{\alpha}}{{\bar{\alpha}}^{(#1)}}}
\newcommand{\alpv}[1]{%
\ifthenelse{\equal{#1}{}}{ \pmb{\alpha}}{{\pmb{\alpha}}_{#1}}}
\newcommand{\alh}[1]{%
\ifthenelse{\equal{#1}{}}{ {\mathbf{a}}}{{{a}}_{#1}}}
\newcommand{\alhh}[1]{%
\ifthenelse{\equal{#1}{}}{ {\hat{\alpv{}}}}{{{\hat{\alpha}}}_{#1}}}

\newcommand{\pauliV}[1]{%
\ifthenelse{\equal{#1}{}}{ {\vec{\pmb{\sigma}}}}{{\pmb{\sigma}}_{#1}}}

\newcommand\Jones{J}
\newcommand\JonesComp{H}
\newcommand\Reeal{R}
\newcommand\Muller{M}
\newcommand\AlgEst{G}

\newcommand\bT[1]{\mathbf{\JonesComp}_{#1}}

\newcommand\bH[1]{\mathbf{\AlgEst}_{#1}}

\newcommand\bR[1]{\mathbf{\Jones}_{#1}}
\newcommand\bRf{\Jones}
\newcommand\bTf{\JonesComp}
\newcommand\bHf{\AlgEst}
\newcommand\bRR[1]{\mathbf{R}_{#1}}
\newcommand\bRRf{\Reeal}
\newcommand{\bb}[1]{%
\ifthenelse{\equal{#1}{}}{ {\vec{\pmb{\rho}}}}{{\pmb{\rho}}_{#1}}}
\newcommand\bbb[1]{
\ifthenelse{\equal{#1}{}}{ {\vec{\pmb{\lambda}}}}{{\pmb{\lambda}}_{#1}}}

\newcommand{\bRl}[1]{%
\ifthenelse{\equal{#1}{}}{\mathbf{R_l}}{\mathbf{R_l}_{#1}}%
}
\newcommand{\bRr}[1]{%
\ifthenelse{\equal{#1}{}}{\mathbf{R_r}}{\mathbf{R_r}_{#1}}%
}

\newcommand{\bRri}[1]{%
\ifthenelse{\equal{#1}{}}{\mathbf{R}_\mathrm{R}^{-1}}{{\mathbf{R}_\mathrm{R}^{(#1)}}^{-1}}%
}
\newcommand{\bRli}[1]{%
\ifthenelse{\equal{#1}{}}{\mathbf{R}_\mathrm{L}^{-1}}{{\mathbf{R}_\mathrm{L}^{(#1)}}^{-1}}%
}

\newcommand\bRRhi[1]{\hat{\bRR{}}_{#1}^{-1}}
\newcommand\bRRh[1]{\hat{\bRR{}}_{#1}}

\newcommand{\bRin}[1]{%
\ifthenelse{\equal{#1}{}}{{\bRf(\innov{\alpv{}})}}{{\bRf(\innov{\alpv{}}_{#1})}}}
\newcommand{\bTin}[1]{%
\ifthenelse{\equal{#1}{}}{{\bTf( \btin{k}, \innov{\alpv{}})}}{{\bTf(\btin{k}, \innov{\alpv{}}_{#1})}}}
\newcommand{\bHin}[1]{%
\ifthenelse{\equal{#1}{}}{{\bHf(\innov{\alpv{}})}}{{\bHf(\innov{\alpv{}}_{#1})}}}
\newcommand{\bRRin}[1]{%
\ifthenelse{\equal{#1}{}}{\bRRf(\innov{\alpv{}})}{\bRRf(\btin{k}, \innov{\alpv{}}_{#1})}}
\newcommand{\bRrin}[1]{%
\ifthenelse{\equal{#1}{}}{\innov{\bRr{}}}{\innov{\bRr{}}_{#1}}}
\newcommand{\bRlin}[1]{%
\ifthenelse{\equal{#1}{}}{\innov{\bRl{}}}{\innov{\bRl{}}_{#1}}}

\newcommand{\pn}[1]{%
\ifthenelse{\equal{#1}{}}{e^{i\phi}}{e^{i\phi_{#1}}}}
\newcommand{\pnin}[1]{%
\ifthenelse{\equal{#1}{}}{e^{i\innov{\phi}}}{e^{i\innov{\phi}_{#1}}}}

\newcommand{\pnm}[1]{%
\ifthenelse{\equal{#1}{}}{e^{-i\phi}}{e^{-i\phi_{#1}}}}
\newcommand{\pnmin}[1]{%
\ifthenelse{\equal{#1}{}}{e^{-i\innov{\phi}}}{e^{-i\innov{\phi}_{#1}}}}

\newcommand{\bt}[1]{%
\ifthenelse{\equal{#1}{}}{\phi}{\phi_{#1}}}
\newcommand{\btin}[1]{%
\ifthenelse{\equal{#1}{}}{\innov{\phi}}{\innov{\phi}_{#1}}}

\newcommand\bM[1]{\mathbf{\Muller}_{#1}}
\newcommand\bMf{\Muller}
\newcommand{\bMin}[1]{%
\ifthenelse{\equal{#1}{}}{{\bMf({\innov{\alpv{}}})}}{{\bMf(\innov{\alpv{}}_{#1})}}}
\newcommand\bMhi[1]{\hat{\bM{}}_{#1}^{-1}}
\newcommand\bMh[1]{\hat{\bM{}}_{#1}}

\newcommand\bS[1]{\mathbf{s}_{#1}}
\newcommand\bSh[1]{\hat{\mathbf{s}}_{#1}}
\newcommand\bV[1]{\mathbf{v}_{#1}}
\newcommand\bVh[1]{\hat{\mathbf{v}}_{#1}}

\newcommand{\autocorr}[2]{%
\ifthenelse{\equal{#1}{}}{\mathcal{A}_{#2}}{\mathcal{A}_{#2}{#1}}%
}
\newcommand{\autocorrinv}[2]{%
\ifthenelse{\equal{#1}{}}{\mathcal{A}^{-1}_{#2}}{\mathcal{A}^{-1}_{#2}{#1}}%
}

\newcommand{\thet}[2]{%
\ifthenelse{\equal{#1}{}}{\ifthenelse{\equal{#2}{}}{\theta}{\theta^{(#2)}}}{\ifthenelse{\equal{#2}{}}{\theta_{#1}}{\theta^{(#2)}_{#1}}}}

\newcommand\cpo[1]{\mathcal{K}({#1})}

\makeatletter
\newcommand{\sbullet}{%
  \hbox{\fontfamily{lmr}\fontsize{.8\dimexpr(\f@size pt)}{0}\selectfont\textbullet}}

\makeatother

\newcommand\partd[2]{\frac{\partial #1}{\partial #2}}
\newcommand\innov[1]{\accentset{\sbullet}{#1}}

\newcommand{\rmT}{\mathrm{T}} 
\newcommand{\rmH}{\mathrm{H}} 
\newcommand{\rmC}{\mathrm{*}} 

\newcommand\alpvo[1]{%
\ifthenelse{\equal{#1}{}}{\hat{\alpv{}}}{\hat{\alpv{}}_{#1}}}
\newcommand\bto[1]{%
\ifthenelse{\equal{#1}{}}{\bt{}}{\hat{\bt{}}_{#1}}}

\newcommand\bTh[1]{\hat{\bT{}}_{#1}}
\newcommand\bThi[1]{\hat{\bT{}}_{#1}^{-1}}
\newcommand{\df}{\Delta \nu} 
\newcommand{\dpp}{\Delta p} 
\newcommand\bI[1]{\mathbf{I}_{#1}}
\newcommand\stPh{\mu_{\bt{}}}
\newcommand\stSOP{\mu_{\alpv{}}}

\newcommand\muCMA{\mu_\mathrm{CMA}}
\newcommand\muMMA{\mu_\mathrm{MMA}}

\DeclareMathAlphabet{\mathpzc}{OT1}{pzc}{m}{it}

\makeatletter
\newcommand{\subalign}[1]{%
  \vcenter{%
    \Let@ \restore@math@cr \default@tag
    \baselineskip\fontdimen10 \scriptfont\tw@
    \advance\baselineskip\fontdimen12 \scriptfont\tw@
    \lineskip\thr@@\fontdimen8 \scriptfont\thr@@
    \lineskiplimit\lineskip
    \ialign{\hfil$\m@th\scriptstyle##$&$\m@th\scriptstyle{}##$\crcr
      #1\crcr
    }%
  }
}
\makeatother

\makeatletter
\newcommand{\raisemath}[1]{\mathpalette{\raisem@th{#1}}}
\newcommand{\raisem@th}[3]{\raisebox{#1}{$#2#3$}}
\makeatother

\makeatletter
\newcommand{\vast}{\bBigg@{4}}
\newcommand{\Vast}{\bBigg@{5}}
\makeatother

\newcounter{defcounter}
\DeclareMathOperator*{\argmin}{arg\,min}

\usepackage[linesnumbered,ruled,vlined,titlenotnumbered]{algorithm2e}
\newlength\algowd
\def\savewd#1{\setbox0=\hbox{#1\hspace{1in}}\algowd=\wd0\relax#1}
\newcommand\algolines[2]{\savewd{#1}
  \tcp*[f]{\parbox[t]{\dimexpr\algowidth-\algowd}{#2}}}

\makeatletter
\newcommand{\algorithmfootnote}[2][\footnotesize]{%
  \let\old@algocf@finish\@algocf@finish
  \def\@algocf@finish{\old@algocf@finish
    \leavevmode\rlap{\begin{minipage}{\linewidth}
    #1#2
    \end{minipage}}%
  }%
}
\makeatother
\makeatletter
\newcommand{\removelatexerror}{\let\@latex@error\@gobble}
\makeatother

\makeatletter
\renewcommand{\sbullet}{%
  \hbox{\fontfamily{lmr}\fontsize{.35\dimexpr(\f@size pt)}{0}\selectfont\textbullet}}

\makeatother

\newcounter{algoline}

\newcolumntype{L}[1]{>{\raggedright\arraybackslash}p{#1}}
\newcolumntype{C}[1]{>{\centering\arraybackslash}p{#1}}
\newcolumntype{R}[1]{>{\raggedleft\arraybackslash}p{#1}}

\begin{document}

\title{Modulation Format Independent Joint Polarization and Phase Tracking for Coherent Receivers}

\author{Cristian~B.~Czegledi,
        Erik~Agrell,~\IEEEmembership{Senior~Member,~IEEE,}
        Magnus~Karlsson,~\IEEEmembership{Senior~Member,~IEEE,~Fellow,~OSA}, 
         and~Pontus~Johannisson,~\IEEEmembership{~Member,~OSA}
\thanks{
C. B. Czegledi and E. Agrell are with the Department of Signals and Systems, Chalmers University of Technology, SE-412 96 Gothenburg, Sweden (e-mail: czegledi@chalmers.se).

M. Karlsson and P. Johannisson are with the Department of Microtechnology and Nanoscience, Chalmers University of Technology, SE-412 96 Gothenburg, Sweden.

This work was supported by the Swedish research council (VR) and performed within the Fiber Optic Communications Research Center (FORCE) at Chalmers. The simulations were performed on resources at Chalmers Centre for Computational Science and Engineering (C3SE) provided by the Swedish National Infrastructure for Computing (SNIC).  
}
}


\maketitle

\begin{abstract}
The state of polarization and the carrier phase drift dynamically during transmission in a random fashion in coherent optical fiber communications. 
The typical digital signal processing solution to mitigate these impairments consists of two separate blocks that track each phenomenon independently.
Such algorithms have been developed without taking into account mathematical models describing the impairments.   
We study a  blind, model-based tracking algorithm to compensate for these impairments. The algorithm dynamically recovers the carrier phase and state of polarization jointly for an arbitrary modulation format.  
Simulation results show the effectiveness of the proposed algorithm, having a fast convergence rate and an excellent tolerance to phase noise and
dynamic drift of the polarization. 
The computational complexity of the algorithm is lower compared to state-of-the-art algorithms at similar or better performance, which makes it a 
strong candidate for future optical systems.
\end{abstract}


\ifCLASSOPTIONpeerreview
 \begin{center} \bfseries EDICS Category: 3-BBND \end{center}
\fi

\IEEEpeerreviewmaketitle

\section{Introduction} 
\IEEEPARstart{D}{igital} signal processing (DSP) enables  spectrally efficient communications based on coherent transmission.  Contrary to  traditional optical transmission links that are based on intensity-modulation and direct-detection, coherent transmissions carry the information in both the intensity and phase of the optical field, in both polarizations, and benefit from improved sensitivities, higher-order modulation formats, and digital  impairment mitigation.
Polarization-multiplexed quadrature phase-shift keying (PM-QPSK) introduced for 100 Gb/s transmission has been widely deployed and reached maturity.  
Recently, 200 Gb/s transceivers have been made commercially available based on $16$-ary polarization-multiplexed quadrature amplitude modulation (PM-16-QAM)
and it is expected that in  the near future, higher-order PM-$M$-QAM modulation formats   will become a necessity for higher data rates. However, the improved spectral efficiency comes at the cost of a reduced tolerance to impairments such as laser phase noise and drift of the state of polarization (SOP), which have to be dynamically tracked in the receiver \cite{Kim2009,Savory2010}.  

The phase and SOP  tracking are important DSP blocks at the receiver and are different from the chromatic dispersion compensation, which can be set once and then forgotten due to its static behavior. The  SOP drift has its origin in the imperfections of the manufacturing process of the fiber cables, mechanical/thermal stress on the deployed  fibers, splices, etc. Due to these random variations, the SOP changes dynamically in time and along the fiber, which makes it difficult to fully compensate for. The phase noise originates from the finite coherence length of the transmitter and receiver lasers and it drifts in time as a Wiener random walk. Despite the fact that the SOP drift and the phase noise arise due to different hardware imperfections, they can  be  modeled jointly as dynamic rotations of the optical field \cite{Czegledi2015a}.
A deterministic or static behavior of these phenomena would be straightforward to resolve, but when the impairments drift randomly, the receiver must adjust dynamically to track the phenomena.

The common DSP solution for SOP tracking is done in the Jones space using the constant modulus algorithm (CMA) \cite{Savory2010}, initially developed for  two-dimensional  modulation formats \cite{Godard1980}, or modified versions of it to accommodate for various modulation formats,  such as the multiple modulus algorithm (MMA)\cite{JianYang2002,Louchet2008} or the polarization-switched \mbox{(PS-)CMA} \cite{Johannisson2011}. Alternatively, the polarization demultiplexing  can be done in the Stokes space \cite{Visintin2014,Muga2014}, which in addition also aligns the phase of the two polarizations, thus improving the phase tracking by enabling joint phase estimation over the two polarizations.   In general, the phase tracking is performed independently of the SOP tracking, using algorithms such as the Viterbi--Viterbi algorithm \cite{Viterbi1983} or the blind phase-search algorithm (BPS) \cite{Pfau2009}, which treat each polarization independently. 

Recently, Louchet \textit{et al.} proposed the {Kabsch} algorithm\cite{Louchet2014}, which addresses the two impairments jointly in both polarizations in the real four-dimensional (4D)  space and  accommodates any modulation format. In general, joint estimation leads to better performance, and it is expected that future transceivers will benefit from  improved performance from such integrations of different DSP blocks\cite{Morero2015}.

However, very few algorithms present in the literature take into consideration analytical  models  describing the behavior of the  impairments. Model-based algorithms have a restricted flexibility and therefore fewer degrees of freedom (DOFs) to adjust. The DOFs of model-based algorithms are restricted to only the ones covered by the impairment to compensate for, thus resulting in a more efficient impairment cancellation, rather than scanning over a larger domain in order to find the optimal setup. 

In this paper, we propose a  model-based algorithm to jointly recover  the carrier phase and SOP  for  arbitrary modulation formats. The design of the algorithm   is  based on a channel model (described in Section~\ref{Sec:model})  that can emulate temporal stochastic polarization and phase drifts, and has been successfully validated with data measured on installed fibers\cite{Czegledi2015a}.  The algorithm (described in Section~\ref{sec:alg}) uses a non-data aided decision-directed architecture, hence zero overhead, and operates jointly on both polarizations. 
 The performance of the algorithm is investigated in Section~\ref{sec:results} by comparing it with state-of-the-art algorithms for different modulation formats, whereas the complexity is evaluated in Section~\ref{sec:complexity}. The proposed algorithm performs similarly or better than state-of-the-art algorithms and provides a good trade-off between complexity and performance regardless of the modulation format. High performance and  fast convergence rate  at low complexity,  for any modulation format,  make the algorithm a strong candidate for future elastic optical systems, where the modulation format can be changed dynamically during transmission to accommodate for various channel and network conditions.

The following notation conventions are used throughout the paper: column vectors are denoted by bold lower case (e.g., $\bx{}$) and matrices by  bold upper case (e.g., $\bf{U}$), except a few specific cases, for literature consistency reasons, denoted by small Greek  letters such as the Pauli matrices $\pauliV{i}$,  the 4D basis matrices $\bb{i}\text{, } \bbb{i}$, and the electric field Jones vector $\mathbf{E}$. Transposition is written as $\bx{}^\mathrm{T}$, conjugation as $\bx{}^\rmC$, and conjugate transpose as $\bx{}^\rmH$. The $n\times n$ identity matrix is written as $\mathbf{I}_n$ and the expectation operator as  $\mathbb{E}[\cdot]$. The dot operation $\alpv{}\cdot\pauliV{}$  should be interpreted as a  linear combination of the three matrices forming the tensor $\pauliV{}=(\pauliV{1},\pauliV{2},\pauliV{3})$.  Multiplication of a matrix with the tensor $\pauliV{}$ result in a tensor with element-wise multiplications, e.g., $\textbf{U} \pauliV{} = (\textbf{U}\pauliV{1},\textbf{U}\pauliV{2},\textbf{U}\pauliV{3})$. The absolute value is denoted by $\abs{\cdot}$ and the Euclidean norm  by $\norm{\cdot}$.

\begin{figure*}
  \centering
  \includegraphics{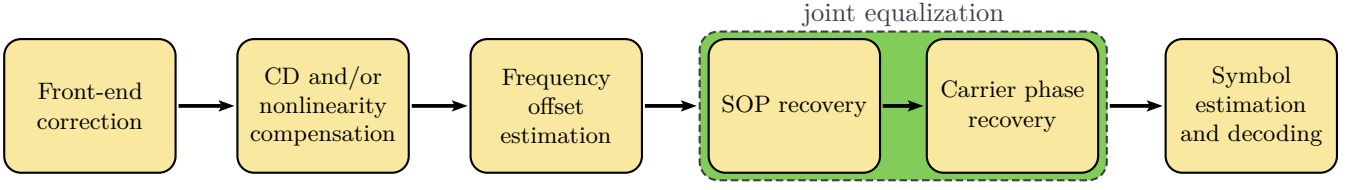}
\caption{Receiver block diagram with elementary DSP modules.}
\label{fig:sysm}
\end{figure*}

\section{Discrete-Time Channel Model} \label{Sec:model}
The coherent optical signal has two quadratures in two polarizations and can be described by a Jones vector
\begin{equation}
  \label{eq:x}
\mathbf{E}(z,t) = 
  \begin{pmatrix}
  E_{x}(z,t) \\
  E_{y}(z,t)
\end{pmatrix},
\end{equation}
 at propagation distance $z$ and time $t$, where   $E_{x}$ ($E_{y}$) is the $x$-polarized ($y$-polarized) electric field, represented as a complex baseband signal. 
The linearly modulated transmitted electric field into the transmission medium  is 
\begin{equation}
  \mathbf{E}(0,t) = \displaystyle \sum_k \bx{k} p(t-kT),
\end{equation}
where $ \bx{k} \in \mathbb{C}^2$ are the information symbols for $k\in\mathbb{Z}$, $T$ is the symbol (baud) interval, and $p(t)$ is a real-valued pulse shape. 

The received discrete symbols, at distance $L$, are obtained from the received electric field $\mathbf{E}(L,t)$ after matched filtering and sampling 
\begin{equation}
\by{k} = \int_{-{\infty}}^{{\infty}} \mathbf{E}(L,t) p^\rmC(t-kT) \mathrm{d} t.
\end{equation}

The discrete transmitted symbols  $\bx{k}$ are drawn independently  from a finite constellation $\mathscr{C} = \{\mathbf{c}_1, \mathbf{c}_2, ..., \mathbf{c}_M\}$ with equal probability. The average energy per symbol is the average of $\norm{\bx{k}}^2$ and  in this case equals
\begin{equation} \label{eq:Es}
  E_s = \frac{1}{M}\sum_{k=1}^M\norm{\mathbf{c}_k}^2.
\end{equation}

Assuming that the chromatic dispersion has been successfully compensated for and polarization-dependent losses and polarization mode dispersion are negligible,  the propagation of the optical field can be described by a unitary $2\times2$ complex-valued Jones matrix $\bR{k}$ \cite[p.~18]{Damask2005}. 
The received symbol $\by{k}\in\mathbb{C}^2$, in the presence of optical amplifier noise, SOP drift, and laser phase noise, can be related  to the input $\bx{k}$ as
\begin{equation}\label{eq:sys_mod}
	\by{k}= \pnm{k}\bR{k} \bx{k}+\bn{k},
\end{equation}
where $i=\sqrt{-1}$,  $\bt{k}$ models the carrier phase noise, and $\bn{k}\in\mathbb{C}^2$ denotes the additive noise, which is represented by two independent complex circular zero-mean Gaussian random variables with variance $N_0/2$  per real dimension, i.e., $\mathbb{E}[\bn{k}\bn{k}^\rmH] = N_0 \bI{2} $\cite{Agrell2009}.

The phase noise is  modeled as a Wiener process \cite{Pfau2009,Tur1985}
\begin{equation}\label{eq:ph_wiener}
  \bt{k+1}=\btin{k}+\bt{k},
\end{equation}
where $\btin{k}$ is the \textit{innovation} of the phase noise. 
The innovation $\btin{k}$ is a random variable drawn independently at each time instance $k$ from a zero-mean Gaussian distribution 
\begin{equation}\label{eq:ph_distr}
  \btin{k} \sim \mathcal{N}(0,\sigma_\nu^2),
\end{equation}
where the variance $\sigma_\nu^2 = 2\pi { \df} T$,  $\df$ is the  sum of the linewidths of the transmitter and receiver lasers. The initial phase  $\bt{0}$ is modeled   as a random variable uniformly distributed in the interval $[0, 2\pi)$.

The  time evolution of the SOP drift can be emulated by modeling $\bR{k}$ as a sequence of random Jones matrices \cite{Czegledi2015a}
\begin{equation} \label{eq:time_dep}
  \bR{k+1} = \bRin{k}\bR{k},
\end{equation}
where $\bRin{k}$ is the SOP \emph{innovation} matrix (cf. $\btin{k}$ in (\ref{eq:ph_wiener})). The matrix function $\bRf(\alpv{}{})$ is defined using the \emph{matrix exponential} \cite[p.~165]{Bellman1960} parameterized by three DOFs $\alpv{}{}$ \cite{Gordon2000} as 
\begin{IEEEeqnarray}{rCl}
\bRf(\alpv{}{}) &=& \exp(-i\alpv{}\cdot\pauliV{}), \label{eq_rot1}
\end{IEEEeqnarray}
where $\alpv{}=(\alp{1}{}, \alp{2}{},\alp{3}{})$ is a three-component real vector and $\pauliV{}=(\pauliV{1},\pauliV{2},\pauliV{3})$ is a tensor of the Pauli spin matrices \cite[eq.~(2.5.19)]{Damask2005}
\begin{equation} \label{eq:pauliSpins}
\pauliV{1} = 
\begin{pmatrix}
  1 & 0 \\
  0&-1\\
\end{pmatrix},
\enspace
\pauliV{2} = 
\begin{pmatrix}
  0 & 1 \\
 1&0\\
\end{pmatrix},
\enspace
\pauliV{3} = 
\begin{pmatrix}
  0 & -i \\
  i &  0\\
\end{pmatrix}.
\end{equation}
The vector $\alpv{}$ can be expressed  as a product \mbox{$\alpv{}=\thet{}{}\alh{}$} of its length $\thet{}{} = \norm*{\alpv{}}$  and the  unit vector $\alh{} = (\alh{1}{}, \alh{2}{},\alh{3}{})$, which represents its direction on the unit sphere. Based on this decomposition of $\alpv{}$, \eqref{eq_rot1} can be rewritten into an explicit form
\begin{IEEEeqnarray}{rCl} \label{eq:J_cos}
\bRf(\alpv{}{})  &=& \bI{2} \cos\thet{}{} - i \alh{}\cdot\pauliV{}\sin\thet{}{}. \label{eq_rot2}
\end{IEEEeqnarray}

Since the transformation $\bRf(\alpv{}{})$ is unitary, the inverse can be found by the conjugate transpose operation or negating the argument, $\bRf(\alpv{}{})^{-1} = \bRf(\alpv{}{})^\rmH=  \bRf(-\alpv{})$.

The random nature of the SOP drift is emulated by drawing the  three innovation parameters $\alpi{}_k$  of the innovation $\bRin{k}$  independently from a zero-mean real Gaussian distribution at each time instance $k$
\begin{equation}\label{eq:rand_alp}
  \alpi{}_k \sim \mathcal{N}(\mathbf{0},\sigma_p^2\, \bI{3}),
\end{equation}
where $\sigma_p^2= 2\pi { \Delta p}\, T$ and  $\Delta p$ is referred as the {\it polarization linewidth}, which quantifies the speed of the SOP drift, analogously to the linewidth describing the phase noise, cf.  (\ref{eq:ph_distr}).

The initial state of the channel $\bR{0} = \bRf(\alpv{0}{})$ is formed from the vector $\alpv{0}= \thet{}{}\alh{}$, which is identified from  the unit vector $(\cos \thet{}{}, \alh{1}{}\sin \thet{}{}, \alh{2}{} \sin \thet{}{},\alh{3}{}\sin \thet{}{})^\rmT = \mathbf{g}/\norm{\mathbf{g}} $ where $\mathbf{g} \sim \mathcal{N}(\mathbf{0}, \bI{4})$. This ensures that $\bR{0} \bx{}$ is uniformly distributed over all possible SOP for a fixed $\bx{}$ \cite{Czegledi2015a,Karlsson2015}.

The phase noise and the SOP drift can be combined into a single operation $\bT{k}=\pnm{k}\bR{k}$ and  (\ref{eq:sys_mod}) can be rewritten as 
\begin{equation}\label{eq:sys_mod_ful}
	\by{k}=  \bT{k} \bx{k}+\bn{k}.
\end{equation}
The update of $ \bT{k}$ can be expressed analogously to (\ref{eq:time_dep}) as 
\begin{equation} \label{eq:T_time_dep}
  \bT{k+1} = \bTin{k}\bT{k},
\end{equation}
where the phase innovation $\btin{k}$ and the random vector $\alpi{}_k$ are defined as (\ref{eq:ph_distr}) and  (\ref{eq:rand_alp}), respectively. The matrix  function $\bTf(\bt{}, \alpv{}{})$ can be expressed as
\begin{IEEEeqnarray}{rCl}
\bTf(\bt{}, \alpv{}{}) &=& \pnm{}\bRf(\alpv{}{}) =  \exp(-i(\alpv{}\cdot\pauliV{}+\bt{}\bI{2})) \label{eq_rot_full}\\
                 &=& ( \cos\bt{} - i \sin\bt{})(\bI{2} \cos\thet{}{} - i \alh{}\cdot\pauliV{}\sin\thet{}{}),\IEEEeqnarraynumspace \label{eq_rot_full2}
\end{IEEEeqnarray}
which combines the effects of both phase noise and SOP drift.

\section{Polarization and Phase Tracking Algorithm} \label{sec:alg}
In order to successfully decode the data at the receiver, the channel matrix $\bT{k}$ (or, equivalently, $\bt{k}$ and $\bR{k}$) needs to be estimated and tracked during transmission such that it is possible to accurately estimate $\bx{k}$ from the received sample $\by{k}$. This section provides a description of the proposed algorithm to estimate $\bT{k}$, first using the Jones formalism, thereafter alternatives using the Stokes and real 4D formalisms are given.

The Jones description can be replaced by the Stokes or real 4D descriptions, which can provide benefits in  some situations \cite{Czegledi2015a}. The analytics describing  wave propagation based on the Jones formalism rely on complex two-dimensional vectors and matrices that have \emph{four} DOFs. This description is sufficient for wave propagation since it can cover any linear phenomenon that can arise during photon propagation. The Stokes description is preferred in some situations since the Stokes vectors  are observable quantities and can be visualized as points on a  three-dimensional sphere, called the Poincar\'e sphere. In this case, the channel matrix $\bR{k}$ is replaced by a $3\times 3$ Mueller matrix $\bM{k}$ with \emph{three} DOFs that models only the changes of the SOP. The Stokes description cannot model absolute phase shifts, therefore it is immune to phase noise.
The real 4D formalism models the channel behavior using $4\times 4$ real rotation matrices that have \emph{six} DOFs, which span over a richer space than 
the  Jones (four DOFs) or Mueller  (three DOFs) matrices can. However, only four of them are physically realizable for propagating photons and the other two can be synthesized using DSP \cite{Karlsson2014}.

In the remainder of this section, we will provide the derivation of the proposed tracking algorithm using the Jones description, after which the equivalent algorithms  using the Stokes and the real 4D descriptions are given. The details of the channel model descriptions using the Stokes and real 4D descriptions are omitted  and can be found in \cite{Czegledi2015a}.
\subsection{Jones Description} \label{sec:Jones_descr}
The considered DSP setup is shown in Fig.~\ref{fig:sysm}, where we combine the SOP drift and carrier phase tracking, i.e., estimation of $\bT{k}$, into a single block after the  frequency offset compensation, which can be done in this case using spectrum-based methods {\cite{Sun2011,Nakagawa2010}}.
Considering that $\bT{k}$ does not change significantly over a symbol duration, we estimate the transmitted symbol from $\bThi{k}\by{k}$ based on a previously calculated estimate of the channel matrix  $\bTh{k}$ using the minimum Euclidean distance criterion
\begin{equation}\label{eq:dec}
  \bxh{k} = \argmin_{\mathbf{c} \in \mathscr{C}} \norm{ { \bThi{k}\by{k}}- \mathbf{c}}^2.
\end{equation}
Thereafter  $\bTh{k}$ is updated as 
\begin{equation}\label{eq:T_hat_up}
  \bTh{k+1} = \bTf(\bto{k}, \alpvo{k})\bTh{k},
\end{equation}
where $\bto{k}$ and  $\alpvo{k}$  are estimates of $\btin{k}  $ and $\alpi{}_k$. These estimates are calculated such that  the error function 
\begin{equation}\label{eq:error_func}
\bbf{k} = \norm{\Big (\bTf(\bt{}, \alpv{}{})\bTh{k}\Big )^{-1} \by{k} - \bxh{k}}^2,
\end{equation}
is  minimized with respect to $\bt{}$ and $\alpv{}{}$, i.e., 
\begin{equation} \label{eq:min_er}
 [\bto{k}, \alpvo{k}] = \argmin_{\bt{}, \, \alpv{}{}} \bbf{k} .
\end{equation}
This can be achieved by computing   $\bto{k}$ and  $\alpvo{k}$  using the   gradient descent method \cite[p.~466]{Boyd2004} 
\begin{IEEEeqnarray}{rCl}
  \bto{k}  &=& - \stPh \mathpzc{Re} \Bigg ( \partd {\bbf{k}}{\bt{}}\bigg\vert_{\raisemath{4.5pt}{{\subalign{\bt{}&=0\\ \alpv{}{}&=[0,0,0]^\rmT}}}} \Bigg )  \label{eq:ph_up} \\
                   &=& -2  \stPh \mathpzc{Re} \Big ( i(\bThi{k} \by{k} - \bxh{k})^\rmH \bThi{k}  \by{k}\Big ), \label{eq:ph_up2}
\end{IEEEeqnarray}
\begin{IEEEeqnarray}{rCl}
  \alpvo{k}  &=& - \stSOP \mathpzc{Re} \Bigg (\nabla_{\alpv{}{}} \bbf{k} \bigg\vert_{\raisemath{4.5pt}{{\subalign{\bt{}&=0\\ \alpv{}{}&=[0,0,0]^\rmT}}}} \Bigg ) \label{eq:SOP_up} \\
                       &=& - 2 \stSOP \mathpzc{Re} \Big ( i(\bThi{k} \by{k} - \bxh{k})^\rmH \bThi{k} \pauliV{} \by{k}\Big ), \label{eq:SOP_up2}
\end{IEEEeqnarray}
where $\stPh$ and $\stSOP$ are positive tracking  step sizes of the phase and of the SOP parameters, respectively, which determine the speed of the algorithm's convergence, the tracking accuracy, and the rate at which changes in the channel can be tracked. The derivations of \eqref{eq:ph_up2}  and \eqref{eq:SOP_up2} can be found in the Appendix. Both innovation parameters $\btin{k}$ and $\alpi{}_k $ have zero mean by (\ref{eq:ph_distr}) and (\ref{eq:rand_alp}); therefore the partial derivatives in (\ref{eq:ph_up}) and (\ref{eq:SOP_up}) are evaluated at $\bt{}=0 , \, \alpv{}{}=[0,0,0]^\rmT$, which results in no preferred direction of $\bto{k}$ and $\alpvo{k}$. Evaluating the gradient at non-zero values could compensate for constant offsets; e.g., using $\bt{} \neq 0$ could compensate for frequency offsets.

It is important to note that $\bTf(\bt{}, \alpv{}{})$ is a many-to-one function. Therefore $\bto{k}$ and $  \alpvo{k}$ are not necessarily equal to $\btin{k}, \alpi{}_k$ and can have different values, but resulting in same matrix $\bTf(\bt{}, \alpv{}{})$.

Since the typical drift time of the SOP  is  slower  ($\sim1~$\si{\ms} or larger) \cite{Ogaki2003, Krummrich2005, Czegledi2015a} than the drift of the phase noise ($\sim1$ \si{\micro\second}) \cite{Pfau2009}, we chose the tracking steps $\stPh, \stSOP$ of the two phenomena differently. For the same reasons, the update of the SOP estimate $\alpvo{k} $ can be done less often than the update of the absolute phase estimate $ \bto{k}$, which will result in a decreased DSP complexity. In this case, $\bto{k}$ can be calculated using (\ref{eq:ph_up2}) at every time instance $k$ and $\alpvo{k}$ using (\ref{eq:SOP_up2}) only at every $P$ symbols, otherwise should be set to $[0,0,0]^\rmT$. 

\begin{figure}[t]
  \centering
  \hspace{0.5cm} 
  \includegraphics{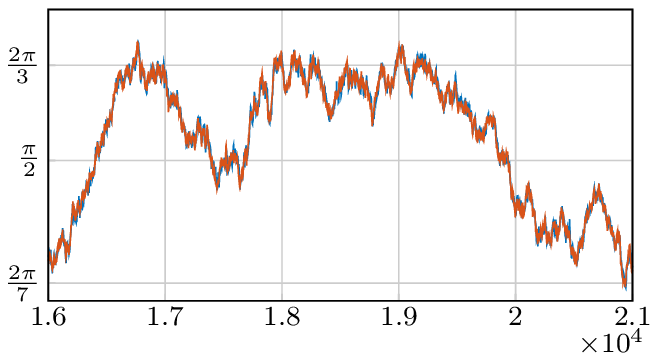}
 \\
 \vspace{0.2cm}
  \includegraphics{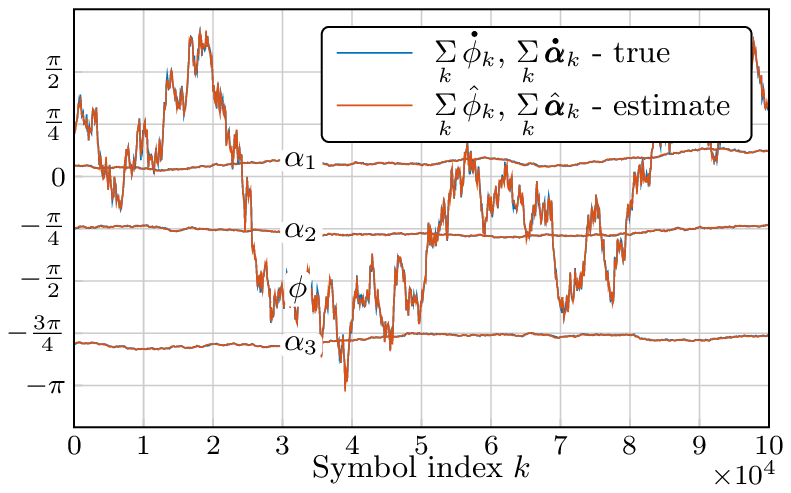}
\psframe[linewidth=0.5pt,framearc=0](-5.85, 4.05)(-6.25, 4.7)
\psline[linewidth=0.5pt](-5.85, 4.7)(-1.3,  5.3)
\psline[linewidth=0.5pt](-6.25, 4.7)(-6.45, 5.3)
\caption{Tracked channel parameters using the proposed algorithm with $\df=1$ MHz and $\dpp=1$ kHz at  28 Gbaud PM-16-QAM are shown. As can be seen,  the algorithm has excellent tracking capabilities without exhibiting cycle slips.}
\label{fig:tracking}
\end{figure}

Fig.~\ref{fig:tracking} shows an example of the algorithm's tracking capability, where we compare the sum of the innovations $\btin{k}, \alpi{}_k$ with their estimates  $\bto{k}, \alpvo{k}$ obtained using the proposed algorithm. Even though, as mentioned above,  $\bto{k}, \alpvo{k}$ do not have to follow $\btin{k}, \alpi{}_k$ to obtain a good estimate of the channel matrix $\bT{k}$, the algorithm manages to obtain similar parameters efficiently without exhibiting cycle slips over the $10^5$ simulated symbols. Perhaps different values of $\bto{k}, \alpvo{k}$ may be obtained if the initial values $\bto{0}, \alpvo{0}$  are not set to be the same as  $\bt{0}, \alpv{0}{}$. Note that the plotted parameters are just for demonstration purposes and do not reflect the behavior of $\bT{k}$ or $\bTh{k}$,  since $\bT{k} \neq \bTf(\underset{\tiny k}{\Sigma} \, \btin{k}, \underset{\tiny k}{\Sigma} \,\alpi{}{}_k)$ because in general $\bTf(\bt{1}+\bt{2}, \alpv{1}{}+\alpv{2}{}) \neq  \bTf(\bt{1}, \alpv{1}{})\bTf(\bt{2}, \alpv{2}{})$.

In a prestudy for this work\cite{Czegledi2015}, we investigated a similar algorithm that tracks both the carrier phase and the SOP jointly. The main difference between the two algorithms consists in the updating rule of $\bTh{k}$, which in (\ref{eq:T_hat_up}) is done as a matrix multiplication. However, the method in \cite{Czegledi2015} was based on a different system model that did not reflect random SOP drift accurately; therefore it is not suited for installed fiber transmissions.

\subsection{Stokes Description}
\label{sec:sts_descr}
In this case, the propagation of the electric field can be  modeled as \cite{Czegledi2015a}
\begin{equation}\label{eq:st_sys_mod}
  \bS{\by{k}} = \bM{k} \bS{\bx{k}} + \bS{\bn{k}},
\end{equation}
where  $\bS{\bx{k}} =\bx{k}^\rmH\pauliV{}\bx{k}$ and   $\bS{\by{k}} =\by{k}^\rmH\pauliV{}\by{k}$ are the corresponding Stokes vectors of $\bx{k}$ and $\by{k}$ \cite[eq.~(2.5.26)]{Damask2005}. The noise term is $\bS{\bn{k}} =  (\bT{k} \bx{k})^\rmH\pauliV{}\bn{k}+\bn{k}^\rmH\pauliV{}\bT{k} \bx{k} + \bn{k}^\rmH\pauliV{}\bn{k}$ and can be identified by equating terms after substituting \eqref{eq:sys_mod_ful} in $\bS{\by{k}} =\by{k}^\rmH\pauliV{}\by{k}$. As can be noted, $\bS{\bn{k}}$ is signal dependent and \eqref{eq:st_sys_mod} is not an additive noise model, opposed to \eqref{eq:sys_mod_ful}.  The channel matrix $\bM{k}$ modeling the evolution of the SOP can be expressed using a $3 \times 3$ Mueller matrix defined as\cite{Gordon2000}
\begin{IEEEeqnarray}{c} 
  \bMf{}(\alpv{}) = \exp(2\cpo{\alpv{}}), \label{eq:st_mat_exp1}
\end{IEEEeqnarray}
where $\cpo{\alpv{}}$ denotes the cross-product operator \cite[eq.~(11)]{Karlsson2014}
\begin{equation}
  \cpo{\alpv{}} =
\begin{pmatrix}
  0 &  -\alpv{3}{} & \alpv{2}{} \\
 \alpv{3}{} &  0 & -\alpv{1}{} \\
  -\alpv{2}{} &  \alpv{1}{} & 0 
\end{pmatrix}.
\end{equation}
The inverse can be obtained as 
$\bMf{}(\alpv{})^{-1}=\bMf{}(\alpv{})^{\rmT}=\bMf(-\alpv{})$. 
The  polarization transformations introduced by $\bM{k}$ can be seen as  {rotations} of the Poincar\'e sphere around the unit vector $\alh{}$ by an angle $2\thet{}{}$.

Analogously to the Jones description, the algorithm decides first which was the transmitted Stokes vector  based on the  minimum Euclidean distance criterion
\begin{equation} \label{eq:st_err_func}
  \bSh{\bx{k}} = \argmin_{\mathbf{c} \in \mathscr{C}} \norm{ { \bMhi{k}\bS{\by{k}}}- \bS{\mathbf{c}}}^2,
\end{equation}
using the inverted estimate of $\bM{k}$, where\footnote{Note that for constellations with rotational symmetry, more than one constellation point  will correspond to the same Stokes vector; e.g.,  the PM-QAM modulation format has a  four-fold rotational symmetry, therefore four distinct constellation points $\mathbf{c}$ will correspond to the same Stokes vector $\bS{\mathbf{c}}$.} $\bS{\mathbf{c}} = \mathbf{c}^\rmH\pauliV{}\mathbf{c}$.  Thereafter, $ \bMh{k}$ is updated analogously to (\ref{eq:T_hat_up}) as
\begin{equation}\label{eq:M_hat_up}
    \bMh{k+1} = \bMf(\alpvo{k}) \bMh{k}.
\end{equation}
Analogously with \eqref{eq:min_er}--\eqref{eq:SOP_up2}, it can be shown that the optimal   $\alpvo{k}=[\alhh{k,1},\alhh{k,2},\alhh{k,3}]^\rmT$ are computed as 
\begin{equation} \label{eq:alpv_M}
 \alhh{k,i} = 4 \stSOP ( \bMhi{k} \bS{\by{k}} - \bSh{\bx{k}})^\rmT \bMhi{k}\cpo{\mathbf{e}_i} \bS{\by{k}}
\end{equation}
for $i = 1, 2 ,3$,  using the gradient descent algorithm such that the Euclidean distance in the Stokes space is minimized. 
The vectors $\mathbf{e}_i$  form the standard basis in $\mathbb{R}^3$.

Note that the error function that minimizes the Euclidean distance is not optimum in this case since $\bS{\bn{k}}$ includes not only noise, but signal--noise interaction. Furthermore, the noise term $\bn{k}^\rmH\pauliV{}\bn{k}$ is non-Gaussian.  Better metrics \cite{Visintin2014} that take into account the non-Gaussian  distribution of the noise can be used, but it is outside the scope of this work.

\subsection{4D Real  Description}
\label{sec:4D-space-descr} 
In the 4D formalism, the phase and SOP drifts are combined and modeled using a $4\times4$ real orthogonal  matrix $\bRR{k}$ \cite{Karlsson2014,Betti1991,Cusani1992}  as
\begin{equation}\label{eq:st_sys_mod}
  \bV{\by{k}} = \bRR{k} \bV{\bx{k}} + \bV{\bn{k}},
\end{equation}
where $\mathbf{v_z}$ for any $\mathbf{z} = [z_1, z_2]^\rmT \in \mathbb{C}^2$ is defined as $[ \mathpzc{Re}(z_1), \mathpzc{Im}(z_1), \mathpzc{Re}(z_2), \mathpzc{Im}(z_2)]^\rmT$.
The channel matrix $\bRR{k}$ can be expressed using the matrix function \cite{Karlsson2014}
\begin{IEEEeqnarray}{rCl}\label{eq:4D_rot}
\bRRf{}(\bt{},\alpv{}) &=& \exp((\bt{},0,0) \cdot \bbb{}-\alpv{}\cdot\bb{}),
\end{IEEEeqnarray}
where $\bb{}=(\bb{1},\bb{2},\bb{3})\text{ and }\bbb{}=(\bbb{1},\bbb{2},\bbb{3})$ are six\footnote{The matrices $\bbb{2}$ and $\bbb{3}$ are not shown since they do not influence $\bRRf{}(\bt{},\alpv{})$ in \eqref{eq:4D_rot}.} constant basis matrices \cite[eqs. (20)--(25)]{Karlsson2014}
\begin{IEEEeqnarray}{rClrCl}
\bb{1} &=& 
\begin{pmatrix}
  0 & -1 &  0 & 0\\
  1 &  0 &  0 & 0\\
  0 &  0 &  0 & 1\\
  0 &  0 & -1 & 0\\
\end{pmatrix},
\enspace&
\bb{2} &=& 
\begin{pmatrix}
  0 &  0 &  0 & -1\\
  0 &  0 &  1 &  0\\
  0 & -1 &  0 &  0\\
  1 &  0 &  0 &  0\\
\end{pmatrix},
\IEEEeqnarraynumspace \\
\bb{3} &=& 
\begin{pmatrix}
  0 &  0 &  1 & 0\\
  0 &  0 &  0 & 1\\
 -1 &  0 &  0 & 0\\
  0 & -1 &  0 & 0\\
\end{pmatrix},
\enspace&
\bbb{1} &=&
\begin{pmatrix}
  0 &  1 &  0 & 0\\
 -1 &  0 &  0 & 0\\
  0 &  0 &  0 & 1\\
  0 &  0 & -1 & 0\\
\end{pmatrix}.
\end{IEEEeqnarray}
The inverse can be obtained as  
$\bRRf{}(\bt{},\alpv{})^{-1}=\bRRf{}(\bt{},\alpv{})^{\rmT}=\bRRf(-\bt{},-\alpv{})$.

Analogously to the Jones/Stokes description, the algorithm decides first which was the transmitted symbol based on the  minimum Euclidean distance criterion
\begin{equation}
  \bVh{\bx{k}} = \argmin_{\mathbf{c} \in \mathscr{C}} \norm{ { \bRRhi{k}\bV{\by{k}}}- \bV{\mathbf{c}}}^2,
\end{equation}
using the inverted estimate of $\bRR{k}$. The estimate of the channel matrix $ \bRRh{k}$ is updated analogously to (\ref{eq:T_hat_up}) as
\begin{equation}\label{eq:4D_time_dep}
  \bRRh{k+1} = \bRRf(\bto{k}, \alpvo{k})\bRRh{k},
\end{equation}
where, analogously with \eqref{eq:ph_up}--\eqref{eq:SOP_up2} and \eqref{eq:alpv_M},
\begin{equation}\label{eq:ph_4D_up}
  \bto{k}  = 2  \stPh  (\hat{\bRR{}}_{k}^{-1} \by{k} - \bxh{k})^\rmT \hat{\bRR{}}_{k}^{-1}  \bbb{1} \by{k},
\end{equation}
\begin{equation} \label{eq:ph_4D_up2}
  \alpvo{k}   = - 2 \stSOP (\hat{\bRR{}}_{k}^{-1} \by{k} - \bxh{k})^\rmT \hat{\bRR{}}_{k}^{-1} \bb{} \, \by{k}.
\end{equation}

In the above description, only four DOFs, i.e., the scalars $\bt{}$, $\alp{1}{}$, $\alp{2}{}$, and $\alp{3}{}$ corresponding to $\bbb{1}$, $\bb{1}$, $\bb{2}$, and $\bb{3}$, respectively,  of the matrix $\bRRf{}(\bt{},\alpv{})$ were used, which correspond to the carrier phase and the SOP drift. The other two DOFs, i.e., the scalars corresponding to $\bbb{2},\bbb{3}$ in (\ref{eq:4D_rot}), can be used to correct certain transmitter and/or receiver hardware imperfections, which cannot be done using Jones or Stokes formalisms, such as \SI{90}{\degree} I/Q error or the time skew between I and Q \cite{Crivelli2014}.

The algorithm presented in this section is fully equivalent to the one in Section \ref{sec:Jones_descr} and will have the same performance, but not to the one in Section \ref{sec:sts_descr}. The latter may have a different performance due to the suboptimal error function and  a separate solution to mitigate the phase noise is required in the latter case since it will not be covered by $\hat{\bM{}}_k$.
\section{Results}\label{sec:results}
We evaluated the achievable performance of the proposed recovery algorithm numerically.  The details of the simulation setup are described in Section~\ref{sec:sim-setup}, whereas in Sections~\labelcref{subsec:pol_tol,subsec:ls_tol,subsec:snr_tol,subsec:conv_rate} various  performance metrics of the algorithm are evaluated. 

\begin{table*}[!t]
		\ra{1.2}
		\caption[]{Algorithm parameters, achievable performance, and hardware complexity}
		\centering
		\begin{tabu}
                  {p{1.4cm}  p{1.6cm}  p{2.45cm} p{1.23cm} p{1.5cm} p{1.27cm} p{1.5cm} C{0.95cm} C{1.2cm}  C{0.8cm}}
                  \toprule
                  & & {Algorithm parameters} & {Max. tol. $\dpp \cdot T$} & \mbox{{Max.~tol. $\dpp$}} \mbox{at 28 Gbaud} & {Max.~tol. $\df \cdot T$} & \mbox{Max. tol. $\df $} \mbox{at 28 Gbaud} &  Operations & Comparisons  & Memory units \\
                  \midrule

                  \multirow{5}{*}{PS-QPSK} 
                  & Kabsch & $N_\mathrm{Kab}=31$     &  \mbox{$ 0.34\cdot 10^{-4}$} & $ 0.95$ MHz  &   $0.91\cdot 10^{-4}$ & $2.55 $ MHz & 132 &  7  & 0.5  \\
                  & \multirow{3}{*}{PS-CMA+BPS} &   $N_{\mathrm{BPS}}=13$     &    \multirow{3}{*}{\mbox{$0.33\cdot 10^{-4}$}}  &    \multirow{3}{*}{$0.93$ MHz} &    \multirow{3}{*}{$6.67\cdot 10^{-4}$}   &    \multirow{3}{*}{$18.68$ MHz} &  \multirow{3}{*}{1320}  &  \multirow{3}{*}{267} &  \multirow{3}{*}{856} \\
                  & & $P_\mathrm{BPS} =32$ \\
                  & & \mbox{$\muCMA=0.04/{E_s}^2$} \\
                  & Proposed~alg. & $c=27$       & \mbox{$  3.20\cdot 10^{-4}$} &  $8.96$ MHz & $11.5\cdot 10^{-4}$ & $32.2$ MHz & 346 & 7 & 8\\

                  \tabucline[0.2pt gray!140]{-}

                  \multirow{5}{*}{PM-QPSK}
                  & Kabsch & $N_\mathrm{Kab}=16$     &\mbox{$   0.17\cdot 10^{-4}$} & $ 0.47$ MHz &   $0.79\cdot 10^{-4 }$ & $ 2.23$ MHz & 171 &   4  & 1  \\
                  & \multirow{3}{*}{CMA+BPS} & $N_{\mathrm{BPS}}=19$   &    \multirow{3}{*}{\mbox{$0.37\cdot 10^{-4}$}}  &    \multirow{3}{*}{$1.04$ MHz} &    \multirow{3}{*}{$6.98\cdot 10^{-4}$}  &    \multirow{3}{*}{$19.54$ MHz} &  \multirow{3}{*}{2936}  &  \multirow{3}{*}{304} &  \multirow{3}{*}{1264} \\
                  & & $P_\mathrm{BPS} =32$ \\
                  & & \mbox{$\muCMA=0.16/{E_s}^2$} \\
                  & Proposed~alg. & $c=64$     &  \mbox{$1.17\cdot 10^{-4}$}& $ 3.28$ MHz &  $ 9.06\cdot 10^{-4}$ & $25.37$ MHz  & 346 & 4   & 8 \\

                  \tabucline[0.2pt gray!140]{-}

                  \multirow{5}{*}{PM-16-QAM}
                  & Kabsch & $N_\mathrm{Kab}=16$    & \mbox{$ 0.44\cdot 10^{-5}$}& $122.1 $ kHz & $  0.15\cdot 10^{-4}$ & $ 0.412$ MHz & 171 &  12  & 1  \\
                  & \multirow{3}{*}{MMA+BPS} & $N_{\mathrm{BPS}}=19$    &    \multirow{3}{*}{\mbox{$0.14\cdot 10^{-5}$}}  &    \multirow{3}{*}{$ 37.8$ kHz} &    \multirow{3}{*}{$1.48\cdot 10^{-4}$}  &    \multirow{3}{*}{$ 4.14$ MHz} &  \multirow{3}{*}{2940}  &  \multirow{3}{*}{840} & \multirow{3}{*}{1264} \\
                  & & $P_\mathrm{BPS} =32$ \\
                  & & \mbox{$\muMMA=0.04/{E_s}^2$} \\
                  & Proposed~alg. & $c=400$    & \mbox{$  2.48\cdot 10^{-5}$}& $694.4 $ kHz  &  $1.35\cdot 10^{-4}$ & $3.78 $ MHz  & 346 & 12  & 8 \\

                  \tabucline[0.2pt gray!140]{-}

                  \multirow{5}{*}{PM-64-QAM}
                  & Kabsch & $N_\mathrm{Kab}=16$   & \mbox{$0.66\cdot 10^{-6}$}& $ 18.4$ kHz &  $ 0.32\cdot 10^{-5}$  & $ 90.2$ kHz & 171 &  28   & 1  \\
                  & \multirow{3}{*}{MMA+BPS} & $N_{\mathrm{BPS}}=19$    &    \multirow{3}{*}{\mbox{$0.11\cdot 10^{-6}$}}  &    \multirow{3}{*}{$ 3.1 $ kHz} &    \multirow{3}{*}{$3.06\cdot 10^{-5}$}  &    \multirow{3}{*}{$856.8 $ kHz} &  \multirow{3}{*}{5768}  &  \multirow{3}{*}{3806} &  \multirow{3}{*}{2480} \\
                  & & $P_\mathrm{BPS} =64$ \\
                  & & \mbox{$\muMMA=0.035/{E_s}^2$} \\
                  & Proposed~alg. & $c=2352$    & \mbox{$4.59\cdot 10^{-6}$}& $128.5 $ kHz  & $  2.54\cdot 10^{-5}$ & $711.2 $ kHz  & 346 & 28 &  8 \\

                  \tabucline[0.2pt gray!140]{-}

                  \multirow{5}{*}{PM-256-QAM}
                  & Kabsch & $N_\mathrm{Kab}=16$    &  \mbox{$0.19\cdot 10^{-6}$}  & $ 5.3$ kHz &  $ 0.74\cdot 10^{-6}$ & $ 20.6$ kHz & 171 &  60  & 1  \\
                  & \multirow{3}{*}{MMA+BPS} & $N_{\mathrm{\mathrm{BPS}}}=19$     &    \multirow{3}{*}{\mbox{$8.42\cdot 10^{-9}$}}  &    \multirow{3}{*}{$ 0.2 $ kHz} &    \multirow{3}{*}{--}  &    \multirow{3}{*}{--} &  \multirow{3}{*}{5814}  &  \multirow{3}{*}{8058} &   \multirow{3}{*}{2480} \\
                  & & $P_\mathrm{BPS} =64$ \\
                  & & \mbox{$\muMMA~= 0.017/{E_s}^2$} \\
                  & Proposed~alg. & $c=6084$     &   \mbox{$1.22\cdot 10^{-6}$} & $ 34.1$ kHz&  $ 6.30\cdot 10^{-6}$ & $ 176.4 $ kHz  & 346 & 60  & 8 \\
                  \bottomrule
		\end{tabu} 
\label{tab:results}
\end{table*}

\subsection{Simulation Setup}
\label{sec:sim-setup}
The considered modulation formats are PS-QPSK, PM-QPSK, PM-16-QAM, PM-64-QAM, and PM-256-QAM at a symbol rate of $28$ Gbaud. The performance is quantified by counting the number of erroneous (4D) symbols to obtain the symbol error rate (SER) at the receiver for various setups in the presence of  laser phase noise,  SOP drift, and additive white Gaussian noise (AWGN). The latter is quantified through the signal-to-noise (SNR) ratio defined as $\mathrm{SNR} = E_s/N_0$. To remove phase ambiguities differential coding \cite[Sec.~2.6.1]{Siemetz2009} was employed independently in each polarization. Note that the presented results when only AWGN is considered (for comparison reasons) still imply differential coding. The presented results are obtained using the Jones description of the algorithm from Section \ref{sec:Jones_descr}.

The proposed algorithm was implemented such that both  \eqref{eq:ph_up2} and \eqref{eq:SOP_up2} were calculated for every $k$, i.e., $P=1$.  The tracking step size $\stPh$ and $\stSOP$ were chosen for each set of system parameters according to the heuristically obtained relations
\begin{equation}\label{eq:miuPh}
\stPh=\frac{\sqrt{\df T c}}{E_s},
\end{equation}
\begin{equation}\label{eq:miuSOP}
\stSOP = \frac{\sqrt{\dpp T c}}{E_s},
\end{equation}
where  $c$ is a constant given in Table~\ref{tab:results}, which depends on the modulation format.  The constant $c$ was optimized to ensure the best performance in the steady-state regime at $\mathrm{SER}=10^{-3}$. In some applications, the linewidth parameters of the optical link may be unknown at the receiver, and therefore it is impossible to accurately compute $\stPh$ and $\stSOP$. To overcome this problem, the $\dpp$ and $\df$ parameters should be overestimated to be on the safe side. Typically, overestimating these parameters does not lead to considerable degradation in the performance.

For comparison, results obtained by the Kabsch algorithm~\cite{Louchet2014} and combinations of the (PS-)CMA/MMA \cite{Johannisson2011, Savory2010, Louchet2008} and BPS \cite{Pfau2009} algorithms are presented. The Kabsch algorithm operates simultaneously on both polarizations in a decision-directed block-wise fashion, using a rectangular filter of size equal to $N_\mathrm{Kab}=16$ \cite{Louchet2014}, except for PS-QPSK, where it was set to $N_\mathrm{Kab}=31$ since we observed better results with a longer block size. The BPS algorithm uses a sliding-window technique in each polarization independently since a relative phase offset between the two polarization may occur. The length of the window was set to $N_\mathrm{BPS}=19$ \cite{Pfau2009}, except for PS-QPSK, where for the same considerations as above the filter length was set to $N_\mathrm{BPS}=13$. The number of test phases of the BPS algorithm was set to $P_\mathrm{BPS} =32$ for PS-QPSK, PM-QPSK, PM-16-QAM and $P_\mathrm{BPS}=64$ for PM-64-QAM and PM-256-QAM \cite{Pfau2009}. The convergence parameter of the \mbox{(PS-)CMA/MMA} algorithm $\muCMA$ was optimized to tolerate the most polarization noise and the chosen values are listed in Table~\ref{tab:results}.

\begin{figure*}
\centering
\hspace{0.6cm} 
\includegraphics[width=0.5\paperwidth]{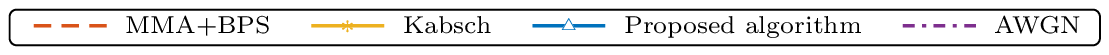}\\
\vspace{0.05cm}
\includegraphics{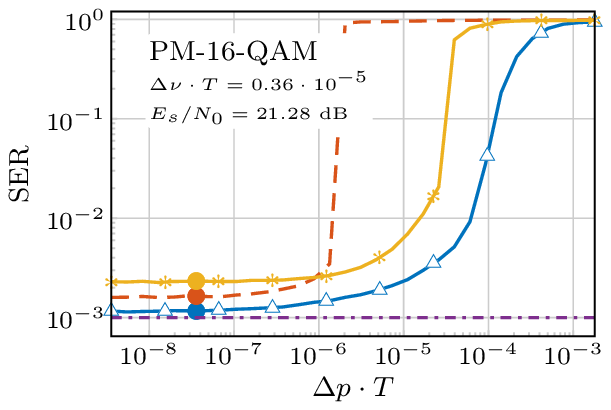}
\includegraphics{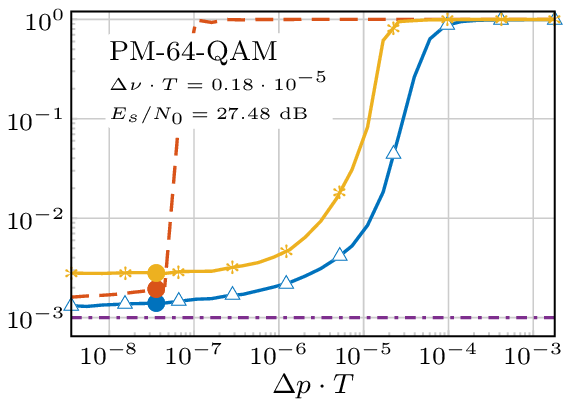}
\includegraphics{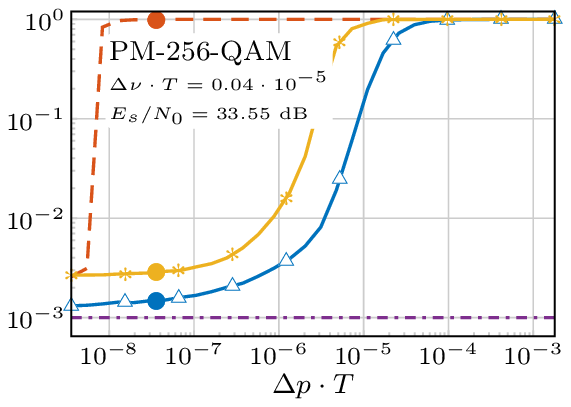}\\
\vspace{0.2cm}
\includegraphics{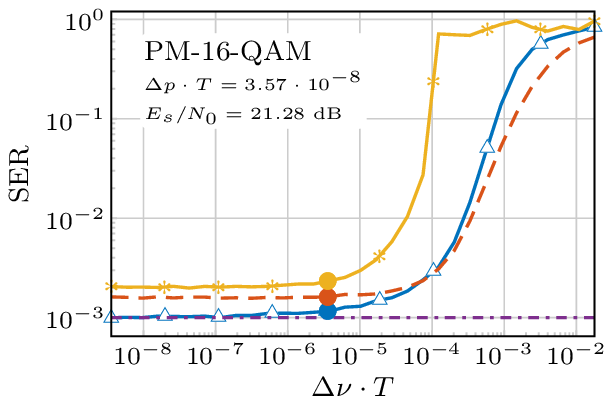}
\includegraphics{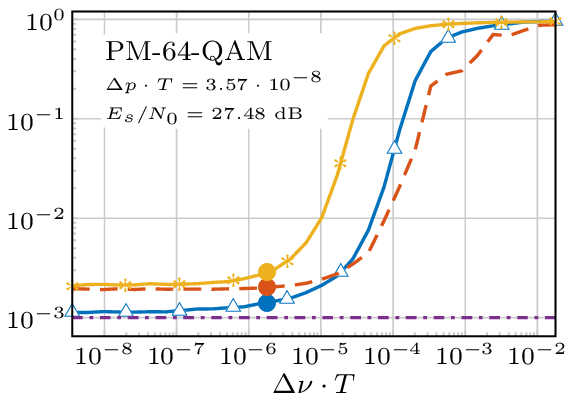}
\includegraphics{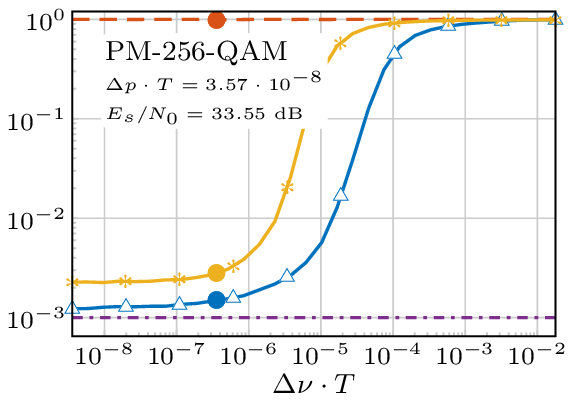}\\
\vspace{0.2cm}
\includegraphics{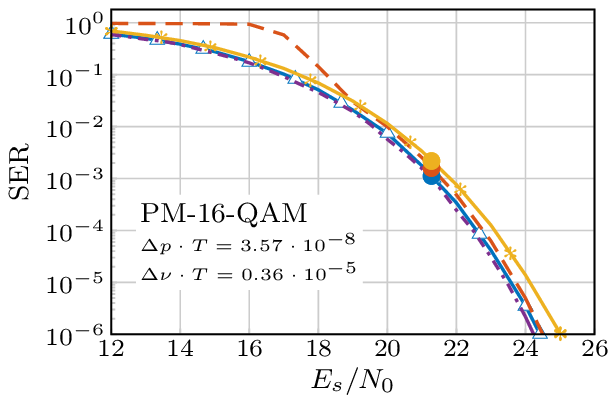}
\includegraphics{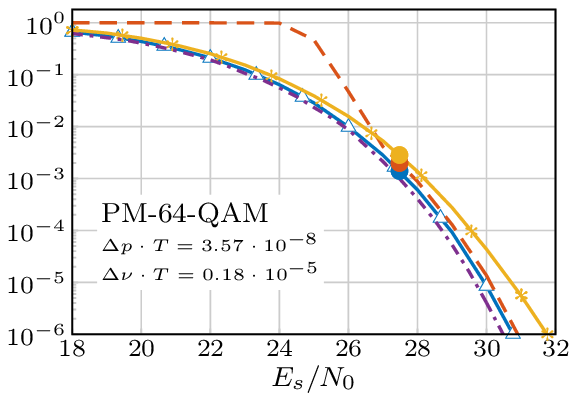}
\includegraphics{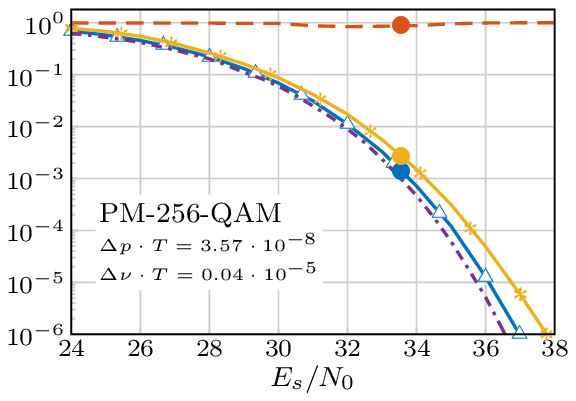}\\
\vspace{0.2cm}
\includegraphics{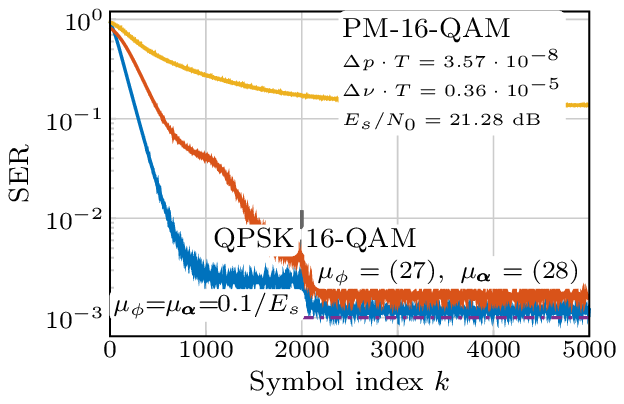}
\includegraphics{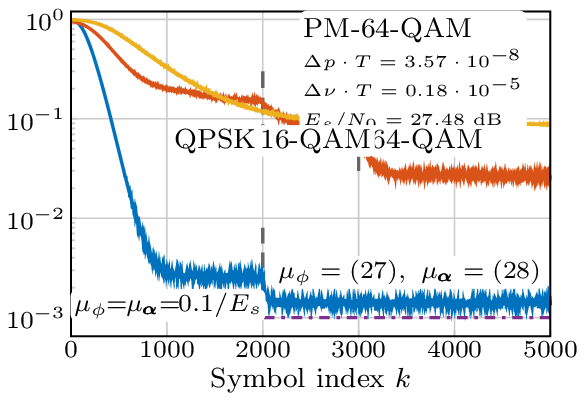}
\includegraphics{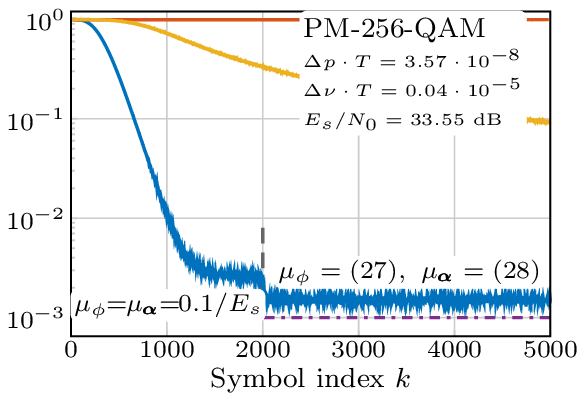}
\caption{The achievable performance of the three tracking schemes for PM-16-QAM, PM-64-QAM, and PM-256-QAM is shown. Each column corresponds to a modulation format, whereas the rows present different performance metrics. The polarization-noise tolerance is shown in the first row by plotting SER versus $\dpp \cdot T$. The penalty compared to the AWGN curve at low $\dpp \cdot T$ is due to the applied phase noise. The tolerance to phase noise is plotted on the second row, where $\df \cdot T$ is varied, whereas the noise sensitivity is shown in the third row by varying the SNR. The round markers shown in the first three rows correspond to the same channel conditions, i.e., the same $\dpp \cdot T$, $\df \cdot T$, and SNR. The convergence  rate is compared on the fourth row, where the SER is plotted versus the symbol index $k$.}
\label{fig:performace}
\end{figure*}

In Sections~\labelcref{subsec:pol_tol,subsec:ls_tol,subsec:snr_tol}, the SOP and phase tracking capabilities of the algorithms are evaluated and therefore the initial channel matrix $\bT{0}$ is considered to be known to the receiver, i.e.,  $\bTh{0}=\bT{0}$. Section~\ref{subsec:conv_rate} presents results on the convergence rate of the algorithms, where the channel matrix $\bT{0}$ is unknown to the receiver. The achieved results are shown in Fig.~\ref{fig:performace}, where each row corresponds to one of  Sections \ref{subsec:pol_tol}--\ref{subsec:conv_rate} and each column corresponds to a modulation format. We omitted results obtained for  PS-QPSK and PM-QPSK in the figure for space reasons,  and these are summarized  in Table~\ref{tab:results}.  

\subsection{Polarization Noise Tolerance} \label{subsec:pol_tol}
In this section, the ability to track time-varying SOP drift is evaluated. To measure the polarization sensitivity, the polarization linewidth $\dpp$ is varied while keeping the other parameters fixed. The SNR is set such that  $\mathrm{SER}=10^{-3}$ is achieved in an AWGN scenario and the accumulated laser linewidth $\df$ is chosen such that $\df\cdot T = \{3.6\cdot 10^{-5}, \, 3.6\cdot 10^{-5}, \,0.36\cdot 10^{-5}, \, 0.18\cdot 10^{-5}, \, 0.04\cdot 10^{-5}\}$ (corresponding to $\df = \{1000, \, 1000, \, 100, \, 50, \, 10\}$ kHz at 28 Gbaud) for PS-QPSK, PM-QPSK, PM-16-QAM, PM-64-QAM, and PM-256-QAM, respectively. 

The results of the simulation are shown in the top row of Fig.~\ref{fig:performace}, where the SER is plotted versus the polarization linewidth times the symbol time.  As can be seen, the proposed algorithm offers the best tolerance to SOP drift at any $\dpp \cdot T$ for all modulation formats. High tolerance to SOP drift enables the use of high-order modulation formats even on rapidly varying channels, such as aerial fibers.
Both competitor algorithms show error floors higher than the proposed algorithm, possibly due to the block-wise operation of the Kabsch algorithm and finite number of test phases of the BPS algorithm, respectively. Perhaps the performance of both  competitor algorithms could be improved if the block size/tracking step/number of test phases would be optimized for each set of system parameters. However, this is outside the scope of the paper. The performance of the MMA-based algorithms   degrades significantly as the size of the constellation increases, providing the worst performance for PM-256-QAM. 

Table \ref{tab:results} summarizes the maximum-tolerable polarization linewidth times the symbol time  such that the receiver requires an extra 1 dB SNR to achieve $\mathrm{SER}=10^{-3}$ compared to the case without laser and polarization noise.  As can be seen in  Table~\ref{tab:results}, the proposed algorithm performs the best in all scenarios, being able to tolerate  up to nine times more polarization noise.

\subsection{Phase-Noise Tolerance} \label{subsec:ls_tol}
Another important feature of the considered tracking algorithms is the tolerance to phase noise. The phase-noise sensitivity is evaluated by varying $\df$ and keeping the rest of the parameters fixed.  The SNR is set such that a $\mathrm{SER}=10^{-3}$ is achieved in an AWGN scenario and the polarization linewidth  is set such that $\dpp\cdot T=3.57 \cdot 10^{-8}$ (corresponding to $\dpp=1$ kHz at 28 Gbaud).

Fig.~\ref{fig:performace}, second row,  shows results obtained by the evaluated algorithms, where the SER is plotted versus the accumulated laser linewidth times the symbol time. The proposed algorithm has similar  performance compared to BPS, which is known to have one of the best phase-noise sensitivities  for QAM modulation formats \cite{Zhou2014}. 

 Table \ref{tab:results} summarizes the maximum tolerable laser linewidth times the symbol time  for an extra 1 dB SNR to achieve $\mathrm{SER}=10^{-3}$ compared to the case without laser and polarization noise. As can be seen in  Table~\ref{tab:results}, the proposed algorithm performs the best in all scenarios, except for PM-16-QAM and PM-64-QAM where it has a slightly worse phase-noise sensitivity, being able to tolerate  laser linewidths  up to $32$ MHz.

\subsection{Additive-Noise Sensitivity} \label{subsec:snr_tol}
Fig.~\ref{fig:performace}, third row, shows the SER as a function of  SNR for fixed $\dpp$ and $\df$ shown in the figure. The Kabsch algorithm has the biggest penalty compared to the AWGN scenario, up $1.2$ dB for PM-256-QAM, while the proposed algorithm has the smallest penalty for all three modulation formats with a maximum of $0.4$ dB for PM-256-QAM. The performance of the MMA+BPS combination  is in between  the two other algorithms, but is the worst at low SNR.

\subsection{Convergence Rate} \label{subsec:conv_rate}
An important quality of non-data aided algorithms is the convergence rate based on blind data.
To compare the convergence rates of the three algorithms, additional numerical simulations have been performed, in which the initial channel matrix $\bT{0}$ was not known by the algorithms  and generated by the method described in Section \ref{Sec:model}. Note that the channel matrix was dynamic and updated according to (\ref{eq:T_time_dep}). In order to compare the algorithms in a fair manner, the initial estimates of the channel matrix were set to be the identity, i.e.,  $\bTh{0}=\bI{2}$ or $\bI{4}$ for the Kabsch algorithm.

The convergence rates  for the different modulation formats are shown in Fig.~\ref{fig:performace}, bottom row, where the SER evolution is plotted versus the symbol index $k$. The results were obtained by averaging over $4\cdot 10^5$ realizations, where in each realization the initial channel matrix was randomly generated.  To improve the convergence rate of the MMA algorithm, the training regime has been split into several stages with increasing numbers of target radii from QPSK to the final constellation \cite{Guiomar2015}, as is shown in the figures. The performance of the Kabsch algorithm could be improved by splitting the training regime into different stages  with different window lengths $N_\mathrm{Kab}$,  but it is outside the scope of this work. The convergence rate of the proposed algorithm is improved by dividing the training regime into two stages, each with different tracking parameters $\stPh, \stSOP$. The first stage, $k=1,\dots,2000$, uses the same value for both step sizes $\stPh=\stSOP=0.1/E_s$, which is bigger than the ones used in the second stage, $k>2000$, where  the values were obtained using (\ref{eq:miuPh}) and (\ref{eq:miuSOP}). It was observed that high step sizes give faster convergence but less accurate tracking in steady state.

Fig.~\ref{fig:performace}, bottom row, shows that the proposed algorithm provides the fastest convergence compared to  the other two algorithms and it approaches the AWGN scenario the closest.  In the considered scenarios, the convergence takes up to $2500$ symbols, depending on the constellation, corresponding to $\sim 100$ {\si{\pico\second}.
 
The convergence rate of the three algorithms can be improved by parallelization \cite{Johannisson2011a}, where the computations are  performed for a number of different initial matrices in parallel and the best candidate is chosen at a later stage.

\section{Hardware Complexity}\label{sec:complexity}
The assessment of the hardware complexity of the three considered algorithms is done by comparing the number of required real\footnote{The complex multiplications have been converted to real operations such that a complex multiplication requires four real multiplications and two real summations, thus six operations in total.}} operations (additions and multiplications), comparisons, and memory units. Although this approach of quantifying complexity does not accurately measure the algorithm's efficiency and  is a rough approximation, it is a starting point.

For pedagogical reasons, in (\ref{eq:T_hat_up}), (\ref{eq:M_hat_up}), and (\ref{eq:4D_time_dep}), the estimate of the channel matrix is computed (instead of the estimate of the inverse), which is then inverted and used throughout the rest of the algorithm. The inversion step is unnecessary  and can be skipped by estimating the inverse of the channel matrix already in (\ref{eq:T_hat_up}),  (\ref{eq:M_hat_up}), and (\ref{eq:4D_time_dep}) by negating $\bto{k}, \alpvo{k}$, thus reducing the required number of operations. In addition, calculations that occurs multiple times in the decoding procedure, e.g., $ \bThi{k}\by{k}$ in \eqref{eq:dec}, \eqref{eq:ph_up2}, \eqref{eq:SOP_up2}, are computed only once.
This holds for all three algorithms. 

In Table~\ref{tab:results},  the obtained results corresponding to one processed 4D symbol are listed. In this evaluation, the required number of operations was minimized by using memory units instead. This choice highly affects the BPS algorithm, where instead of recomputing previously done calculations for every processed symbol, the values are stored in the memory.  Therefore, the results are different from the ones presented in \cite{Ke2012}, where no memory was used.  

As can be seen in Table~\ref{tab:results}, the Kabsch algorithm\footnote{We considered the Golub--Reinsch method \cite[p.~493]{Golub2013} for singular value decomposition required by the  algorithm.} has the lowest complexity since it operates in a block-wise fashion processing the entire block at a time. This reduces the complexity significantly, which depends on the block length, i.e., a longer block will require less computing power per symbol, but also degrades the performance since it assumes that the channel does not change  during the block.  The \mbox{(PS-)CMA/MMA+BPS} combination requires the highest computational effort and it increases with the constellation size and the size of the sliding-window over which the BPS algorithm operates.
The complexity of the proposed algorithm  was evaluated using the Jones description, which is the same with respect to different modulation formats and it is in between the other two algorithms.
The overall complexity of the algorithm can be reduced  if the SOP tracking is not updated at every symbol in (\ref{eq:T_hat_up}), as discussed in Section \ref{sec:Jones_descr}. This is a reasonable approach since the SOP does not drift as fast as the phase noise and it can be considered constant over a limited observation time  in some scenarios. In this case, if $\alpvo{k}$ is computed every $P$ symbols and $\bto{k}$ (which compensates for the phase noise) is updated at every symbol, the number of required operations is reduced from $346$ (Table~\ref{tab:results}) to $203+143/P$. Of course in this scenario the performance will degrade, thus resulting in less tolerable polarization noise, but still sufficient for most installed fiber links.

\section{Algorithmic Summary}
This section  provides an easily implementable form of the proposed algorithm without requiring knowledge about the details of the derivations. 

\removelatexerror
\begin{algorithm}[t]
\label{alg:a}
\caption{Proposed algorithm}
\setcounter{algoline}{-1}
\SetAlgoNlRelativeSize{-0.5}
\KwIn{$\by{k}$, $\bH{k}$, $k$}
\KwOut{$\bxh{k}$, $\bH{k+1}$}
$\bH{0} = \mathbf{I}_2$ \tcp*[f]{\hspace{-0.2cm}initialize the channel matrix} 
\\
{
\algolines{$\bxh{k} = \displaystyle \argmin_{\mathbf{c} \in \mathscr{C}} \norm{ { \bH{k}\by{k}}- \mathbf{c}}^2$ \hspace{0.25cm}}{\hspace{-0.15cm}\mbox{decide the} \mbox{\hspace{0.5cm} symbol}} \\
\If(\tcp*[f]{\hspace{-0.2cm}set the tracking steps}){$k\leq 2000$}{$\stPh=\stSOP=0.1/E_s$ \tcp*[f]{\hspace{-0.2cm}convergence stage}}
\Else{
$
\stPh  =\frac{\sqrt{\df T c}}{E_s}
$ \tcp*[f]{\hspace{-0.2cm}tracking stage}
\\
$
\stSOP = \frac{\sqrt{\dpp T c}}{E_s}
$
}
$
  \bto{k}    = -2  \stPh \mathpzc{Re} \Big ( i(\bH{k} \by{k} - \bxh{k})^\rmH \bH{k}  \by{k}\Big )
$
\\
$
  \alpvo{k}  = - 2 \stSOP \mathpzc{Re} \Big ( i(\bH{k} \by{k} - \bxh{k})^\rmH \bH{k} \pauliV{} \by{k}\Big ) \label{alg:alpv_line}
$
\\
\algolines{$\bH{k+1} =\bH{k}  \bTf(-\bto{k}, -\alpvo{k})$\hspace{0.51cm}}{\hspace{-0.2cm}\mbox{update the} \mbox{\hspace{-1.25cm} channel matrix}}
}
\end{algorithm}

The algorithm is summarized in Algorithm \ref{alg:a}, where for ease of notation we denote $\bH{k}=\bThi{k}$ to be the inverted estimate of the channel matrix at time $k$. The algorithm receives as  inputs the received symbol $\by{k}$, the previous inverse estimate of the channel matrix $\bH{k}$, and the symbol index $k$, and outputs the decided symbol $\bxh{k}$ and the updated  matrix $\bH{k}$.  The step sizes, which should be precomputed, are higher during convergence  ($k\leq2000$) than tracking  ($k>2000$). In the tracking stage,  they are computed   based on the laser linewidth $\df$ and the polarization linewidth $\dpp$, which should be overestimated if they are unknown to the receiver. The Pauli matrices $ \pauliV{}$ are given in \eqref{eq:pauliSpins}, $\mathscr{C}$ is the set of constellation points, ${E_s}$ is the average symbol energy \eqref{eq:Es}, $T$ is the symbol time, and  $\bTf(\cdot)$ is computed according to (\ref{eq_rot_full}). 

To decrease the computational effort, for slowly-varying channels, the update of  $\bH{k}$ can be done less frequently than for every received symbol. Moreover, since the SOP drift varies slower than the phase noise, the SOP update of  $\bH{k}$ can be done less often by not calculating $\alpvo{k}$ in line~\ref{alg:alpv_line} at each iteration but setting it to $\alpvo{k}=[0,0,0]^\rmT$ instead.

The description shown in Algorithm \ref{alg:a} uses the Jones formalism. This can be interchanged with the Stokes or real 4D formalisms by replacing the Jones matrix $\bH{k}$ with $ \hat{\bM{}}_k^{-1}$ (Section~\ref{sec:sts_descr}) or $\hat{\bRR{}}_k^{-1}$ (Section~\ref{sec:4D-space-descr}), and $\bto{k}$, $\alpvo{k}$ should be calculated using (\ref{eq:alpv_M}) or \eqref{eq:ph_4D_up}--\eqref{eq:ph_4D_up2}, respectively. In this case, the Jones vectors $\by{k}$, $\bxh{k}$ become Stokes vectors or real 4D vectors. The Jones description of the algorithm is fully equivalent with the real 4D description and they achieve the same performance. However, the latter involves more computations since multiplying two  $4\times 4$ real matrices requires more operations than multiplying two  $2\times 2$ complex matrices. Nevertheless, the real 4D description can be modified to account for hardware imperfections. On the other hand, the Stokes description is immune to absolute phase shifts and only tracks the SOP, requiring a separate solution for absolute phase tracking. This can be beneficial in some situations since fast oscillations of the phase noise will not affect the algorithm, and therefore the SOP tracking. Performance-wise, the Stokes description will behave differently from the other two as it relies on a suboptimal error function \eqref{eq:st_err_func}.

\section{Discussion and Conclusions}\label{sec:conclusions}
We have proposed a model-based algorithm to jointly track  random polarization and phase drifts. The algorithm uses the gradient descent optimization algorithm in a decision-directed architecture processing one symbol at a time. The achievable performance of the algorithm was evaluated by means of numerical simulations and compared to  state-of-the-art  algorithms. Results show the effectiveness of the proposed algorithm, having a fast convergence rate and an excellent tolerance to phase noise and dynamic drifts of the polarization, in particular when using high-order modulation formats. At similar or better performance, the computational complexity of our algorithm is lower compared to state-of-the-art algorithms. The increased  phase and polarization  noise tolerance for high-order modulation formats at low complexity makes the algorithm a strong candidate  for future  optical systems.

The proposed algorithm is tested in a somewhat idealized scenario assuming that chromatic dispersion, polarization-dependent losses, and polarization mode dispersion are nonexistent. In general, the polarization demultiplexing is performed using finite impulse response filters in conjunction with the CMA or MMA algorithms, which, in addition to SOP recovery,  perform channel equalization.  This equalization is not covered by the proposed algorithm, but  it can be overcame by  coupling the algorithm with a separate equalization stage as in  \cite{Buchali2015} to compensate for the slowly-varying transfer characteristics of the channel.

\appendix\label{app:grad}
In this appendix, the gradient of the error function (\ref{eq:error_func}) with respect to $\bt{}, \alpv{}{}$ is derived as
\begin{IEEEeqnarray}{rCl}
\nabla_{\alpv{}{}}  \bbf{k}  &=& \nabla_{\alpv{}{}}  \norm{\Big (\bTf(\bt{}, \alpv{}{})\bTh{k}\Big )^{-1} \by{k} - \bxh{k}}^2 \label{eq:Er_dev1}\\
&=& 
\mathpzc{Re} \Bigg( 2 \bigg ( \Big (\bTf(\bt{}, \alpv{}{})\bTh{k}\Big )^{-1} \by{k} - \bxh{k} \bigg )^\rmH \nonumber \\
& & 
\qquad \cdot  \nabla_{\alpv{}{}}  \bigg ( \Big (\bTf(\bt{}, \alpv{}{})\bTh{k}\Big )^{-1} \by{k} - \bxh{k} \bigg )  \Bigg) \label{eq:Er_dev2}\\
&=& 
2 \mathpzc{Re} \Bigg(  \bigg ( \bThi{k} \bTf(-\bt{}, -\alpv{}{}) \by{k} - \bxh{k} \bigg )^\rmH \bThi{k} \nonumber  \\
& & 
\qquad \cdot  \nabla_{\alpv{}{}}  \bTf(-\bt{}, -\alpv{}{}) \by{k} \Bigg),\label{eq:Er_dev4}
\end{IEEEeqnarray}
where \eqref{eq:Er_dev2} follows because, for any $ \mathbf{y} \in \mathbb{C}^n$ and  $ \mathbf{x} \in \mathbb{R}^m$, $\nabla_\mathbf{x}\norm{\mathbf{y} }^2= \nabla_\mathbf{x} (\mathbf{y}^\rmH \mathbf{y}) =2\mathpzc{Re}(\mathbf{y}^\rmH \nabla_\mathbf{x} \mathbf{y})$.
The partial derivatives of $ \bTf(-\bt{}, -\alpv{}{}) = \pn{}\bRf(-\alpv{}{})$ are, from \eqref{eq:J_cos} and \eqref{eq_rot_full},
\begin{IEEEeqnarray}{rCl}
   \partd{ \bTf(-\bt{}, -\alpv{}{}) }{\bt{}} &=&  i\pn{}\bRf(-\alpv{}{}),  \label{eq:T_dev3}  \\
   \partd{ \bTf(-\bt{}, -\alpv{}{}) }{ \alp{i}{}} \label{eq:TT_dev1}
&=&
\pn{} \partd{ \bRf(-\alpv{}{}) }{ \alp{i}{}}  \label{eq:TT_dev2}\\
&=&
\pn{} \partd{\Big (\bI{2} \cos\thet{}{}  + i \alh{}\cdot\pauliV{}\sin\thet{}{}\Big )  }{ \alp{i}{}}  \label{eq:TT_dev3}\\
&=&
\pn{}   \bigg(-\bI{2}\alh{i}\sin\thet{}{} +  i   \Big (  \frac{\pauliV{i}}{\thet{}{}}- \frac{\alh{i}}{\thet{}{}} \alh{}\cdot\pauliV{} \Big) \sin \thet{}{}\nonumber
\\
&&  \hspace{1.5cm} 
+ i \alh{}\cdot\pauliV{} \, \alh{i}\cos \thet{}{}
 \bigg ) , \IEEEeqnarraynumspace\label{eq:TT_dev4}
\end{IEEEeqnarray}
where  \eqref{eq:TT_dev4} follows because $\partial \thet{}{} / \partial \alp{i}{} = \alp{i}{} /  \norm*{\alpv{}} = \alh{i}$.

Evaluating  (\ref{eq:T_dev3}) and (\ref{eq:TT_dev4}) at $\bt{}=0 , \, \alpv{}{}=[0,0,0]^\rmT$ results in 
\begin{equation} \label{eq:part-d}
     \partd{ \bTf(-\bt{}, -\alpv{}{}) }{\bt{}} = i \bI{2} , \qquad \quad      \partd{ \bTf(-\bt{}, -\alpv{}{}) }{ \alp{i}{}}  = i\pauliV{i},
\end{equation}
as  $\thet{}{} \to 0$ for any direction $\alh{}$.
The gradient of the error function can be obtained by substituting \eqref{eq:part-d} and $\bTf(0,0)=\bI{2}$ in  (\ref{eq:Er_dev4}), which is then substituted in \eqref{eq:ph_up} and \eqref{eq:SOP_up} to obtain \eqref{eq:ph_up2}  and \eqref{eq:SOP_up2}, respectively.

\ifCLASSOPTIONcaptionsoff
  \newpage
\fi

\bibliographystyle{IEEEtran}


\begin{thebibliography}{10}
\providecommand{\url}[1]{#1}
\csname url@samestyle\endcsname
\providecommand{\newblock}{\relax}
\providecommand{\bibinfo}[2]{#2}
\providecommand{\BIBentrySTDinterwordspacing}{\spaceskip=0pt\relax}
\providecommand{\BIBentryALTinterwordstretchfactor}{4}
\providecommand{\BIBentryALTinterwordspacing}{\spaceskip=\fontdimen2\font plus
\BIBentryALTinterwordstretchfactor\fontdimen3\font minus
  \fontdimen4\font\relax}
\providecommand{\BIBforeignlanguage}[2]{{%
\expandafter\ifx\csname l@#1\endcsname\relax
\typeout{** WARNING: IEEEtran.bst: No hyphenation pattern has been}%
\typeout{** loaded for the language `#1'. Using the pattern for}%
\typeout{** the default language instead.}%
\else
\language=\csname l@#1\endcsname
\fi
#2}}
\providecommand{\BIBdecl}{\relax}
\BIBdecl

\bibitem{Kim2009}
R.~Kim, M.~O'Sullivan, {Kuang-Tsan Wu}, {Han Sun}, A.~Awadalla, D.~Krause, and
  C.~Laperle, ``{Performance of dual-polarization {QPSK} for optical transport
  systems},'' \emph{Journal of Lightwave Technology}, vol.~27, no.~16, pp.
  3546--3559, Aug. 2009.

\bibitem{Savory2010}
S.~J. Savory, ``{Digital coherent optical receivers: algorithms and
  subsystems},'' \emph{IEEE Journal of Selected Topics in Quantum Electronics},
  vol.~16, no.~5, pp. 1164--1179, Sept.--Oct. 2010.

\bibitem{Czegledi2015a}
C.~B. Czegledi, M.~Karlsson, E.~Agrell, and P.~Johannisson, ``Polarization
  drift channel model for coherent fibre-optic systems,'' \emph{Nature
  Scientific Reports}, to appear, 2016.

\bibitem{Godard1980}
D.~N. Godard, ``{Self-recovering equalization and carrier tracking in
  two-dimensional data communication systems},'' \emph{IEEE Transactions on
  Communications}, vol.~28, no.~11, pp. 1867--1875, Nov. 1980.

\bibitem{JianYang2002}
J.~Yang, J.-J. Werner, and G.~A. Dumont, ``{The multimodulus blind equalization
  and its generalized algorithms},'' \emph{IEEE Journal on Selected Areas in
  Communications}, vol.~20, no.~5, pp. 997--1015, Jun. 2002.

\bibitem{Louchet2008}
H.~Louchet, K.~Kuzmin, and A.~Richter, ``{Improved {DSP} algorithms for
  coherent 16-{QAM} transmission},'' in \emph{Proc. European Conference on
  Optical Communication (ECOC)}, Brussels, Belgium, Sept. 2008, p. Tu.1.E.6.

\bibitem{Johannisson2011}
P.~Johannisson, M.~Sj\"{o}din, M.~Karlsson, H.~Wymeersch, E.~Agrell, and P.~A.
  Andrekson, ``{Modified constant modulus algorithm for polarization-switched
  {QPSK}},'' \emph{Optics Express}, vol.~19, no.~8, pp. 7734--7741, Apr. 2011.

\bibitem{Visintin2014}
M.~Visintin, G.~Bosco, P.~Poggiolini, and F.~Forghieri, ``{Adaptive digital
  equalization in optical coherent receivers with Stokes-space update
  algorithm},'' \emph{Journal of Lightwave Technology}, vol.~32, no.~24, pp.
  4759--4767, Dec. 2014.

\bibitem{Muga2014}
N.~J. Muga and A.~N. Pinto, ``{Adaptive 3-D Stokes space-based polarization
  demultiplexing algorithm},'' \emph{Journal of Lightwave Technology}, vol.~32,
  no.~19, pp. 3290--3298, Oct. 2014.

\bibitem{Viterbi1983}
A.~Viterbi and A.~Viterbi, ``{Nonlinear estimation of {PSK}-modulated carrier
  phase with application to burst digital transmission},'' \emph{IEEE
  Transactions on Information Theory}, vol.~29, no.~4, pp. 543--551, Jul. 1983.

\bibitem{Pfau2009}
T.~Pfau, S.~Hoffmann, and R.~No\'e, ``{Hardware-efficient coherent digital
  receiver concept with feedforward carrier recovery for \textit{M}-{QAM}
  constellations},'' \emph{Journal of Lightwave Technology}, vol.~27, no.~8,
  pp. 989--999, Apr. 2009.

\bibitem{Louchet2014}
H.~Louchet, K.~Kuzmin, and A.~Richter, ``{Joint carrier-phase and polarization
  rotation recovery for arbitrary signal constellations},'' \emph{IEEE
  Photonics Technology Letters}, vol.~26, no.~9, pp. 922--924, May 2014.

\bibitem{Morero2015}
D.~A. Morero, M.~A. Castrill\'on, A.~Aguirre, M.~R. Hueda, and O.~E. Agazzi,
  ``{Design trade-offs and challenges in practical coherent optical transceiver
  implementations},'' \emph{Journal of Lightwave Technology}, to appear, 2016.

\bibitem{Damask2005}
J.~N. Damask, \emph{{Polarization Optics in Telecommunications}}.\hskip 1em
  plus 0.5em minus 0.4em\relax New York, NY: Springer, 2005.

\bibitem{Agrell2009}
E.~Agrell and M.~Karlsson, ``{Power-efficient modulation formats in coherent
  transmission systems},'' \emph{Journal of Lightwave Technology}, vol.~27,
  no.~22, pp. 5115--5126, Nov. 2009.

\bibitem{Tur1985}
M.~Tur, B.~Moslehi, and J.~Goodman, ``{Theory of laser phase noise in
  recirculating fiber-optic delay lines},'' \emph{Journal of Lightwave
  Technology}, vol.~3, no.~1, pp. 20--31, Feb. 1985.

\bibitem{Bellman1960}
R.~Bellman, \emph{Introduction to Matrix Analysis}.\hskip 1em plus 0.5em minus
  0.4em\relax New York, NY: McGraw-Hill, 1960.

\bibitem{Gordon2000}
J.~P. Gordon and H.~Kogelnik, ``{{PMD} fundamentals: polarization mode
  dispersion in optical fibers},'' \emph{Proceedings of the National Academy of
  Sciences of the United States of America}, vol.~97, no.~9, pp. 4541--4550,
  Apr. 2000.

\bibitem{Karlsson2015}
M.~Karlsson, C.~B. Czegledi, and E.~Agrell, ``{Coherent transmission channels
  as 4d rotations},'' in \emph{Proc. Signal Processing in Photonic
  Communication (SPPcom)}, Boston, MA, Jul. 2015, p. SpM3E.2.

\bibitem{Karlsson2014}
M.~Karlsson, ``{Four-dimensional rotations in coherent optical
  communications},'' \emph{Journal of Lightwave Technology}, vol.~32, no.~6,
  pp. 1246--1257, Mar. 2014.

\bibitem{Sun2011}
H.~Sun and K.-T. Wu, ``{A novel dispersion and {PMD} tolerant clock phase
  detector for coherent transmission systems},'' in \emph{Proc. Optical Fiber
  Communication Conference (OFC)}, Los Angeles, CA, Mar. 2011, p. OMJ4.

\bibitem{Nakagawa2010}
T.~Nakagawa, K.~Ishihara, T.~Kobayashi, R.~Kudo, M.~Matsui, Y.~Takatori, and
  M.~Mizoguchi, ``{Wide-range and fast-tracking frequency offset estimator for
  optical coherent receivers},'' in \emph{Proc. European Conference on Optical
  Communication (ECOC)}, Torino, Italy, Sept. 2010, p. We.7.A.2.

\bibitem{Boyd2004}
S.~Boyd and L.~Vandenberghe, \emph{Convex Optimization}.\hskip 1em plus 0.5em
  minus 0.4em\relax New York, NY: Cambridge University Press, 2004.

\bibitem{Ogaki2003}
K.~Ogaki, M.~Nakada, Y.~Nagao, and K.~Nishijima, ``{Fluctuation differences in
  the principal states of polarization in aerial and buried cables},'' in
  \emph{Proc. Optical Fiber Communication Conference (OFC)}, Atlanta, GA, Mar.
  2003, p. MF13.

\bibitem{Krummrich2005}
P.~Krummrich, E.-D. Schmidt, W.~Weiershausen, and A.~Mattheus, ``{Field trial
  results on statistics of fast polarization changes in long haul WDM
  transmission systems},'' in \emph{Proc. Optical Fiber Communication
  Conference (OFC)}, Anaheim, CA, Mar. 2005, p. OThT6.

\bibitem{Czegledi2015}
C.~B. Czegledi, E.~Agrell, and M.~Karlsson, ``{Symbol-by-symbol joint
  polarization and phase tracking in coherent receivers},'' in \emph{Proc.
  Optical Fiber Communication Conference (OFC)}, Los Angeles, CA, Mar. 2015, p.
  W1E.3.

\bibitem{Betti1991}
S.~Betti, F.~Curti, G.~{De Marchis}, and E.~Iannone, ``{A novel multilevel
  coherent optical system: 4-quadrature signaling},'' \emph{Journal of
  Lightwave Technology}, vol.~9, no.~4, pp. 514--523, Apr. 1991.

\bibitem{Cusani1992}
R.~Cusani, E.~Iannone, A.~M. Salonico, and M.~Todaro, ``{An efficient
  multilevel coherent optical system: {M-4Q-QAM}},'' \emph{Journal of Lightwave
  Technology}, vol.~10, no.~6, pp. 777--786, Jun. 1992.

\bibitem{Crivelli2014}
D.~E. Crivelli, M.~R. Hueda, H.~S. Carrer, M.~{del Barco}, R.~R. L\'opez,
  P.~Gianni, J.~Finochietto, N.~Swenson, P.~Voois, and O.~E. Agazzi,
  ``{Architecture of a single-chip 50 {G}b/s {DP-QPSK/BPSK} transceiver with
  electronic dispersion compensation for coherent optical channels},''
  \emph{IEEE Transactions on Circuits and Systems I: Regular Papers}, vol.~61,
  no.~4, pp. 1012--1025, Apr. 2014.

\bibitem{Siemetz2009}
M.~Seimetz, \emph{{High-Order Modulation for Optical Fiber
  Transmission}}.\hskip 1em plus 0.5em minus 0.4em\relax Heidelberg, Germany:
  Springer, 2009.

\bibitem{Zhou2014}
X.~Zhou, ``{Efficient clock and carrier recovery algorithms for single-carrier
  coherent optical systems: a systematic review on challenges and recent
  progress},'' \emph{IEEE Signal Processing Magazine}, vol.~31, no.~2, pp.
  35--45, Mar. 2014.

\bibitem{Guiomar2015}
F.~P. Guiomar, S.~B. Amado, A.~Carena, G.~Bosco, A.~Nespola, A.~Teixeira, and
  A.~N. Pinto, ``{Fully-blind linear and nonlinear equalization for {100G
  PM-64QAM} optical systems},'' \emph{Journal of Lightwave Technology},
  vol.~33, no.~7, pp. 1265--1274, Apr. 2015.

\bibitem{Johannisson2011a}
P.~Johannisson, H.~Wymeersch, M.~Sj\"{o}din, A.~S. Tan, E.~Agrell, P.~A.
  Andrekson, and M.~Karlsson, ``{Convergence comparison of the {CMA} and {ICA}
  for blind polarization demultiplexing},'' \emph{Journal of Optical
  Communications and Networking}, vol.~3, no.~6, pp. 493--501, Jun. 2011.

\bibitem{Ke2012}
J.~H. Ke, K.~P. Zhong, Y.~Gao, J.~C. Cartledge, A.~S. Karar, and M.~A. Rezania,
  ``{Linewidth-tolerant and low-complexity two-stage carrier phase estimation
  for dual-polarization 16-{QAM} coherent optical fiber communications},''
  \emph{Journal of Lightwave Technology}, vol.~30, no.~24, pp. 3987--3992, Dec.
  2012.

\bibitem{Golub2013}
G.~H. Golub and C.~F. {Van Loan}, \emph{{Matrix Computations}}, 4th~ed.\hskip
  1em plus 0.5em minus 0.4em\relax Baltimore, MD: Johns Hopkins Univ. Press,
  2013.

\bibitem{Buchali2015}
F.~Buchali, H.~B\"ulow, K.~Schuh, and W.~Idler, ``{{4D-CMA}: enabling
  separation of channel compensation and polarization demultiplex},'' in
  \emph{Proc. Optical Fiber Communication Conference (OFC)}, Los Angeles, CA,
  Mar. 2015, p. Th2A.15.

\end{thebibliography}

\end{document}